\documentclass[12pt,trackchanges,preprint]{aastex63}
\usepackage{amssymb,amsbsy,amsmath}
\usepackage{verbatim,placeins}
\usepackage{graphicx}\usepackage{graphicx} 

\begin{document}

\title{High-fidelity 3D Reconstruction of Solar Coronal Physics with the Updated CROBAR Method}
\correspondingauthor{Joseph Plowman}
\email{jplowman@boulder.swri.edu}

\author[0000-0001-7016-7226]{Joseph Plowman}
\affil{Southwest Research Institute,
Boulder, CO 80302 USA}

\author[0000-0002-0494-2025]{Daniel B. Seaton}
\affil{Southwest Research Institute,
Boulder, CO 80302 USA}

\author[0000-0001-8702-8273]{Amir Caspi}
\affil{Southwest Research Institute,
Boulder, CO 80302 USA}

\author[0000-0003-3410-7650]{J. Marcus Hughes}
\affil{Southwest Research Institute,
Boulder, CO 80302 USA}

\author[0000-0002-0631-2393]{Matthew J. West}
\affil{Southwest Research Institute,
Boulder, CO 80302 USA}

\begin{abstract}
We present an extension of the Coronal Reconstruction Onto B-Aligned Regions (CROBAR) method to Linear Force Free Field (LFFF) extrapolations, and apply it to the reconstruction of a set of AIA, MDI, and STEREO EUVI data. The results demonstrate that CROBAR can not only reconstruct coronal emission structures, but also that it can help constrain the coronal field extrapolations via the LFFF's helicity $\alpha$ parameter. They also provide a real-world example of how CROBAR can easily incorporate information from multiple perspectives to improve its reconstructions, and we also use the additional perspectives to help validate the reconstructions. We furthermore touch on the use of real-world emission passbands rather than idealized power-law type ones using DEMs. We conclude with a comparison of CROBAR generated emission to observed emission and those produced with idealized DEM based power-laws. These results further illustrate the promise of CROBAR for real-world applications, and we make available a preliminary release of the software available for download.
\end{abstract}

\section{Introduction}
\label{sec:intro}
The solar corona is composed of roughly million-degree plasma threaded by a strong (relative to the plasma pressure) magnetic field. It is a unique observational environment for large-scale magnetohydrodynamic phenomena -- the range of temperature and density scales are inaccessible to laboratory experiments or computer simulations, hence the need for and promise of learning about these processes from solar observations. 

Processes in the corona are volumetric (as opposed to occurring on a surface or thin layer) and observations of the corona -- such as extreme ultraviolet (EUV) observations from the Atmospheric Imaging Assembly \citep[AIA;][]{Lemen2012} on board the Solar Dynamics Observatory \citep[SDO;][]{sdo_paper} -- are similarly volumetric. However, the observations are also line-of-sight-integrated as an intrinsic part of the optical remote-sensing process. The line-of-sight confusion caused by this integration impedes our ability to measure the state of the plasma at a given point within the volume: we seek a 3D cube of measurements but can only observe a 2D line-of-sight down-projection of that cube, from, at most, a handful (and often only one) of vantage points.

This is a follow-up paper on a data-driven modeling method called `Coronal Reconstruction Onto B-Aligned Regions' \citep[CROBAR;][]{CROBAR_2021,CROBAR_MURaM} that can resolve this line-of-sight integration ambiguity. CROBAR does so by taking advantage of the fact that the magnetic field that gives the lower corona its structure must follow a set of differential equations (Maxwell's Equations) that constrain the field's geometry (e.g., it must be divergenceless). These constraints, in turn, reduce the number of degrees of freedom that the coronal plasma can have from a fully general 3D cube to something that can be reproduced from a small number of 2D images. Previous papers \citep{CROBAR_2021,CROBAR_MURaM} have laid out the basic details, shown an initial example, and demonstrated that the B-aligned reconstruction is capable of recovering emission properties given a magnetic skeleton.

In this paper, we demonstrate an extension of the magnetic field extrapolation component from the potential field case (no currents in the volume) to the linear force-free field (LFFF) case (currents must be proportional to and aligned with the field). This allows information about the emission observations to be fed back into the underlying magnetic field reconstruction and yields better overall recovery of the field and plasma configuration. We also demonstrate how it is straightforward to incorporate multiple perspectives into the reconstruction using both STEREO and AIA observations.

\section{Linear Force-Free Field with Application}
\label{sec:LFFF_with_application}
The initial paper on CROBAR \citep{CROBAR_2021} used the `potential' magnetic field approximation. This potential approximation assumes that the current everywhere in the coronal volume of interest is negligible. It is the simplest non-trivial approximation for the magnetic field. The only currents are outside the volume or on the boundaries. As a result, the magnetic field is a solution to Laplace's equation, mathematically identical to the electrostatic problem with a distribution of point charges. The solution can be written as an integral over the magnetic charge density $\rho_B$, using the inverse square law as a Green's function for the charge at each point:
\begin{equation}\label{eq:laplace}
    \mathbf{B}(\mathbf{r}) = \int \rho_B(\mathbf{r}')\frac{(\mathbf{r}-\mathbf{r}')}{|\mathbf{r}-\mathbf{r}'|^{3/2}}d^3\mathbf{r}'
\end{equation}
The magnetic charge density, $\rho_B$, is generally found by requiring that the field match observations (e.g., from magnetographs) at the photospheric boundary.

The Linear Force-Free Field (LFFF) approach is a slightly more complex level of approximation, and assumes that, rather than being zero, the current is both aligned with and proportional to the magnetic field everywhere in the volume of interest, with the same constant of proportionality everywhere (the {\em non}-Linear Force Free Field -- nLFFF -- is the next level of complexity, and allows the constant of proportionality to change from one field line to the next, though it must still be constant along any given field line). In other words, in addition to Maxwell's equations, the LFFF requires
\begin{equation}\label{eq:FFF_equation}
    \mathbf{J}(\mathbf{r}) = \alpha\mathbf{B}(\mathbf{r}),
\end{equation}
with $\alpha$ constant everywhere in the LFFF approximation. 

In Equation~\ref{eq:FFF_equation}, the magnetic force (proportional to $\mathbf{J}\times\mathbf{B}$) on the current-carrying plasma is guaranteed to be zero, hence the name of the approximation. The constant $\alpha$ introduces a helicity to the magnetic field, has units of inverse length, and is proportional to the amount of distance required for the field to make one revolution around a common axis. It should be noted that Force-Free-Field extrapolations become unstable if the field lines are long enough to exceed one revolution (i.e., $l/\alpha > 1$).

There are a variety of ways to write the solutions to the LFFF equations; see \cite{WiegelmannSakuraiLRSP} for examples. One way is in terms of a modified version of the Green's function expansion given above. In this work we use the version derived by \cite{ChiuHilton77}. We use a Cartesian coordinate system with the z-axis aligned with the radial vector at the center of the field of view of the image used for the reconstruction, and we assume that the magnetic field $B_0$ on the photospheric boundary is aligned with the z axis (this is the `nominal' case described in \cite{ChiuHilton77}). The Green's-function-based LFFF equations in this coordinate system are:
\begin{equation}\label{eq:LFFF_Green_Bx}
    B_x = \int B_0(x',y') (C_1 \Delta x + \alpha C_2 \Delta y) dx'dy',
\end{equation}
\begin{equation}\label{eq:LFFF_Green_By}
    B_y = \int B_0(x',y') (C_1 \Delta y + \alpha C_2 \Delta x) dx'dy', \mathrm{and}
\end{equation}
\begin{equation}\label{eq:LFFF_Green_Bz}
    B_z = \int B_0(x',y') (\frac{\alpha \Delta z \sin{(\alpha \Delta r)}}{\Delta r^2} + \frac{\Delta z \cos{(\alpha \Delta r)}}{\Delta r^3} ) dx' dy',
\end{equation}
where the two coefficients in $B_x$ and $B_y$ are:
\begin{equation}\label{eq:LFFF_Green_C1}
    C_1 = \left( \alpha\sin{(\alpha \Delta z)} + \frac{\cos{\alpha \Delta r}}{\Delta r} - \frac{\alpha\Delta z \sin{(\alpha\Delta r)}}{\Delta r^2} - \frac{\Delta z^2 \cos{(\alpha \Delta r)}}{\Delta r^3}\right)\frac{1}{\Delta x^2 + \Delta y^2}, \quad \mathrm{and}
\end{equation}
\begin{equation}\label{eq:LFFF_Green_C2}
    C_2 = \frac{\Delta z\cos{(\alpha \Delta r)}}{\Delta r (\Delta x^2 + \Delta y^2)} - \frac{\cos{(\alpha \Delta z)}}{\Delta x^2 + \Delta y^2}.
\end{equation}
Here we have also substituted $\Delta_x = x-x', \Delta_y = y-y', \Delta z = z-z'$, and $\Delta r = \sqrt{\Delta x^2 + \Delta y^2 + \Delta z^2}$. 

\subsection{Curved Coordinates}
\label{sec:curved_coords}
The solar surface is curved, and we want to capture this curvature in CROBAR's reconstructions. On the scale of a small active region (AR), this curvature is not significant. For a large one, however, it is. \cite{ChiuHilton77} mention that they assume a planar photosphere boundary, however, it does not appear that the assumption is actually used. We have numerically tested their equations using a boundary that is not at constant z, and found that the resulting field structure still satisfies the LFFF equations and Maxwell's equations. As best we can tell, the \cite{ChiuHilton77} equations are valid for a curved lower boundary. 

We have continued to assume that the magnetic field at the boundary is along the z axis of the cartesian grid, however. A more accurate approach might be to assume that the field at the photospheric boundary is radial, but this adds considerable complexity to the LFFF equations (off of the `nominal' case) -- the radial vector picks up $x$ and $y$ components further from the center of the field of view. The loss of fidelity that results from this assumption is small, so we leave this improvement to future work.

\subsection{Application of LFFF}
\label{sec:application_of_LFFF}
Here we show the application of the LFFF equations with CROBAR to a solar active region. Later in this paper (see Sec.~\ref{sec:multiple_perspectives}) we will show an example multiple-perspective reconstruction as well. To highlight the improvement that results from a two-perspective reconstruction, we use the same data in both of these sections, with three separate views of the active region to be reconstructed, one or two of which serve(s) as the basis for the reconstruction and the remaining serving as independent view(s), for validation.

Currently the only viable source of off-Sun-Earth-line observational data for such a reconstruction is the Solar TErrestrial RElations Observatory \citep[STEREO][]{STEREO_Kaiser2008} mission. (Solar Orbiter now enables some options as well, but its EUV passbands are more limited for active region observations.) We have identified a narrow window of time at the beginning of the AIA mission where both of the STEREO spacecraft were observing an overlapping region of the Sun that was also visible from Earth, and we use an AR from this interval -- specifically NOAA AR 11089 on July 25, 2010 -- for the demonstrations in this paper. We were unable to download HMI magnetograms for this date from the Virtual Solar Observatory \cite[VSO;][]{VSO2009}, so we use a magnetogram from the Solar and Heliospheric Observatory \citep[SOHO;][]{SOHO_paper} Michelson Doppler Imager \citep[MDI;][]{MDI_paper} instead.

\begin{figure}
    \includegraphics[width=\textwidth]{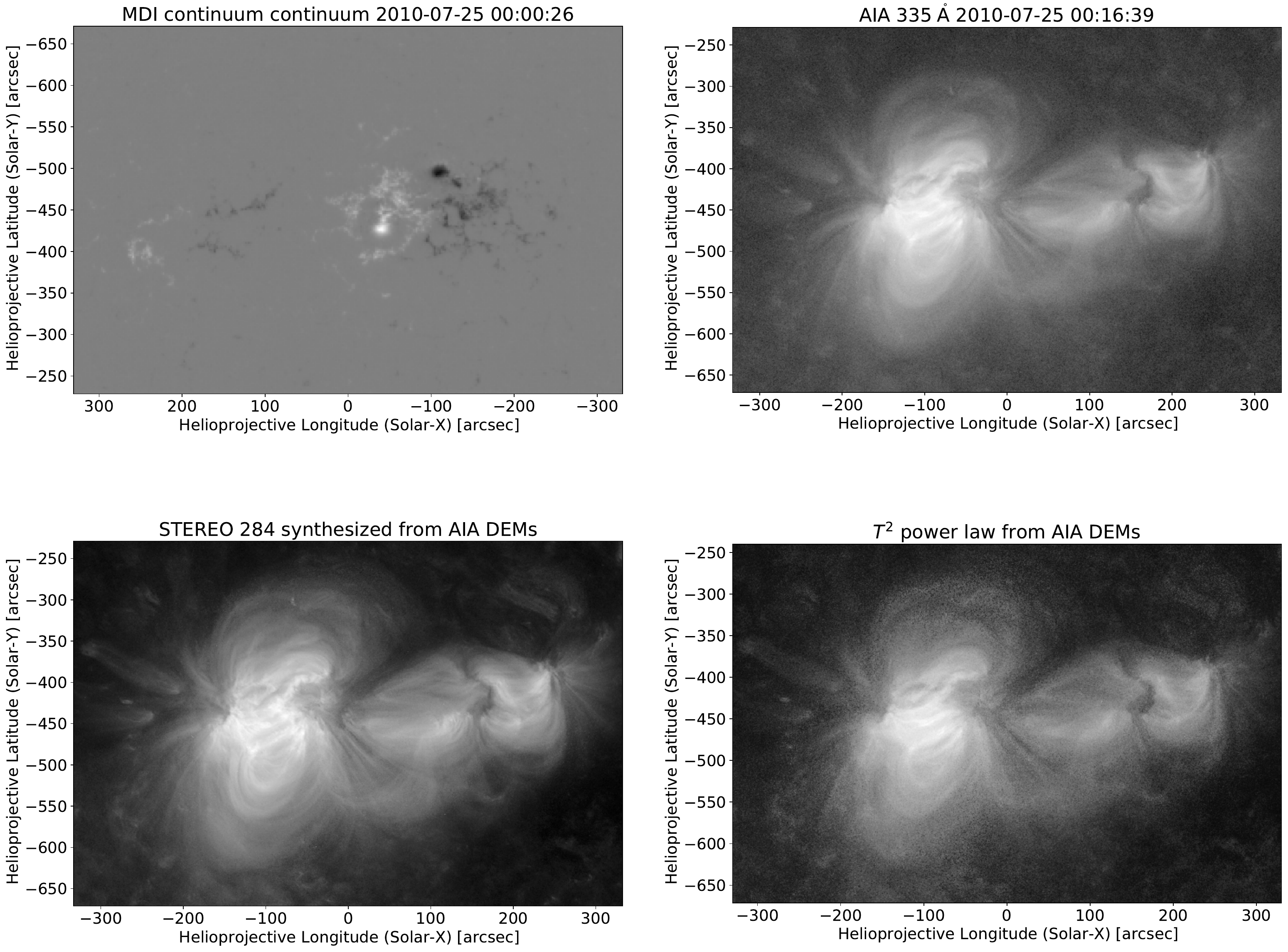}
    \caption{MDI magnetogram (top left) and AIA-derived data used in the reconstruction. Top right is AIA 335\,\AA\ data, lower right is the $T^2$ synthetic temperature response function which is ideal for CROBAR, lower left is STEREO 284\,\AA\ synthesized from AIA DEMs, which will be used in the multi-vantage point reconstruction in conjunction with actual STEREO data.}\label{fig:AR_example_image}
\end{figure}

In \cite{CROBAR_2021}, we show that an ideal passband for CROBAR reconstructions uses a synthetic image made by integrating AIA DEMs against a power law of index 2. We will use this synthetic passband for the initial single-perspective reconstructions (shown first). Figure~\ref{fig:AR_example_image} shows this synthetic passband, alongside an AIA 335\,\AA\ image and the corresponding MDI magnetogram. We also show an image of the STEREO 284\,\AA\ temperature response synthesized from the AIA DEMs, since we will be using that later on in the multi-vantage-point reconstructions with STEREO data.

\begin{figure}
    \includegraphics[width=0.33\textwidth]{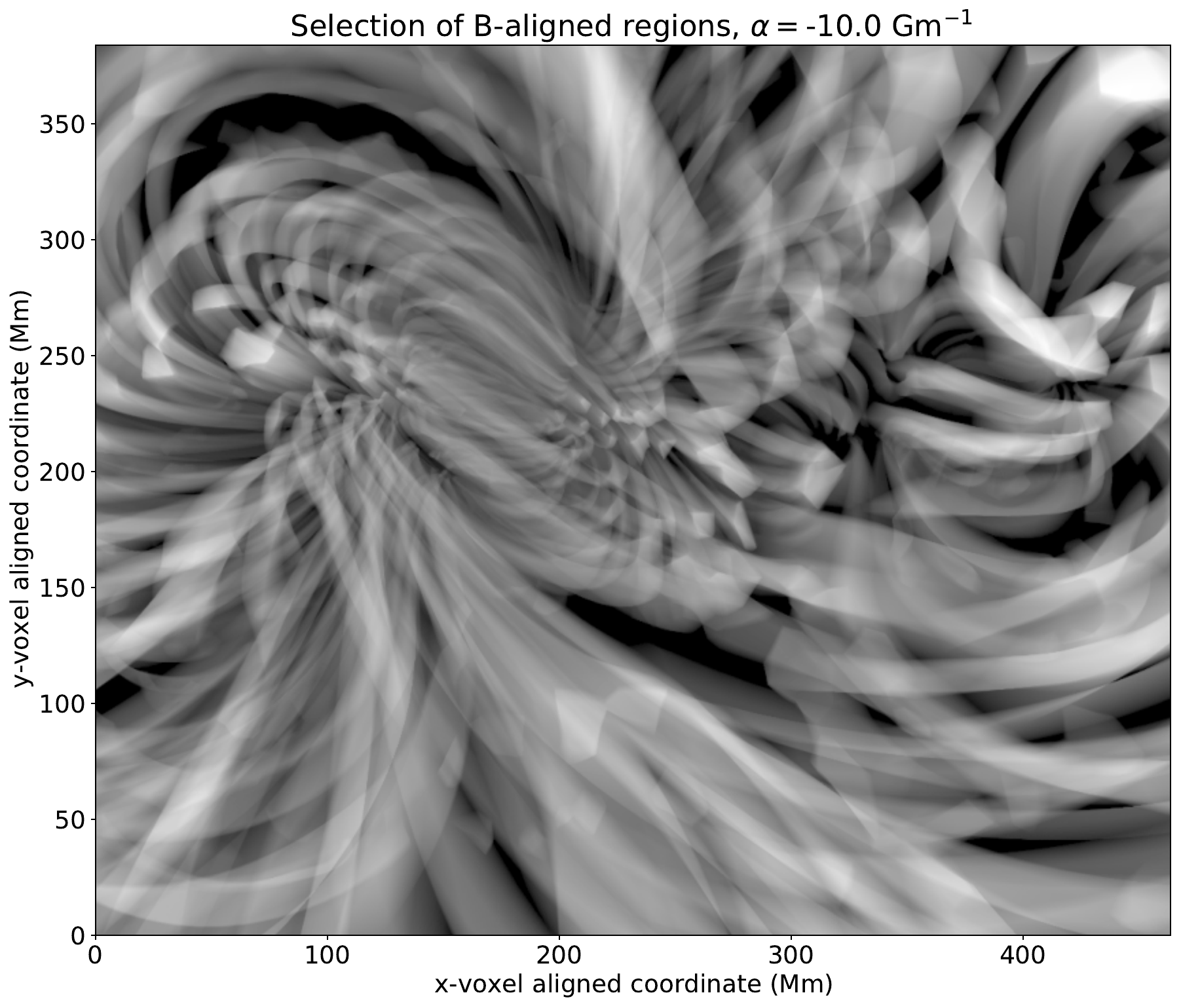}\includegraphics[width=0.33\textwidth]{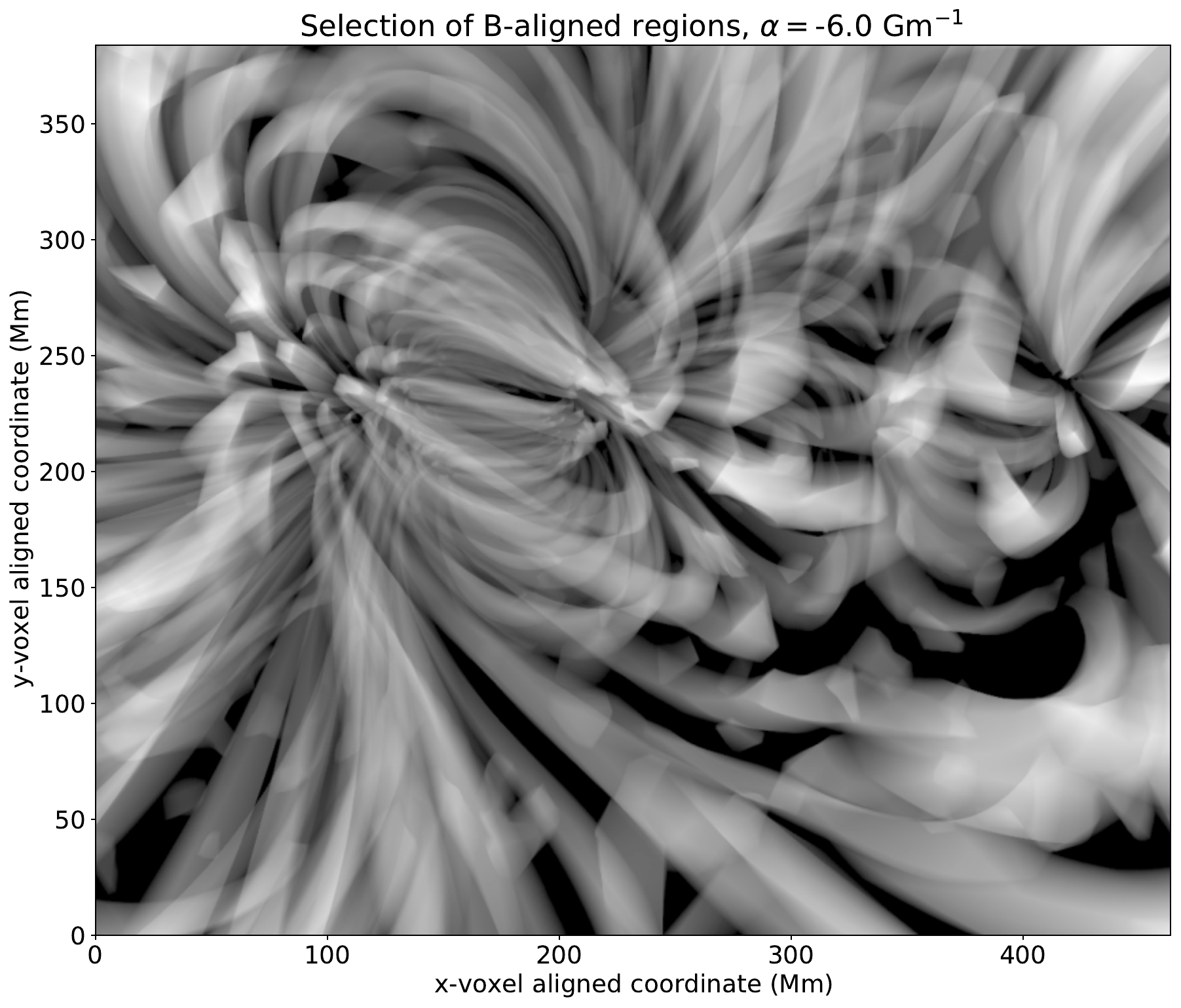}\includegraphics[width=0.33\textwidth]{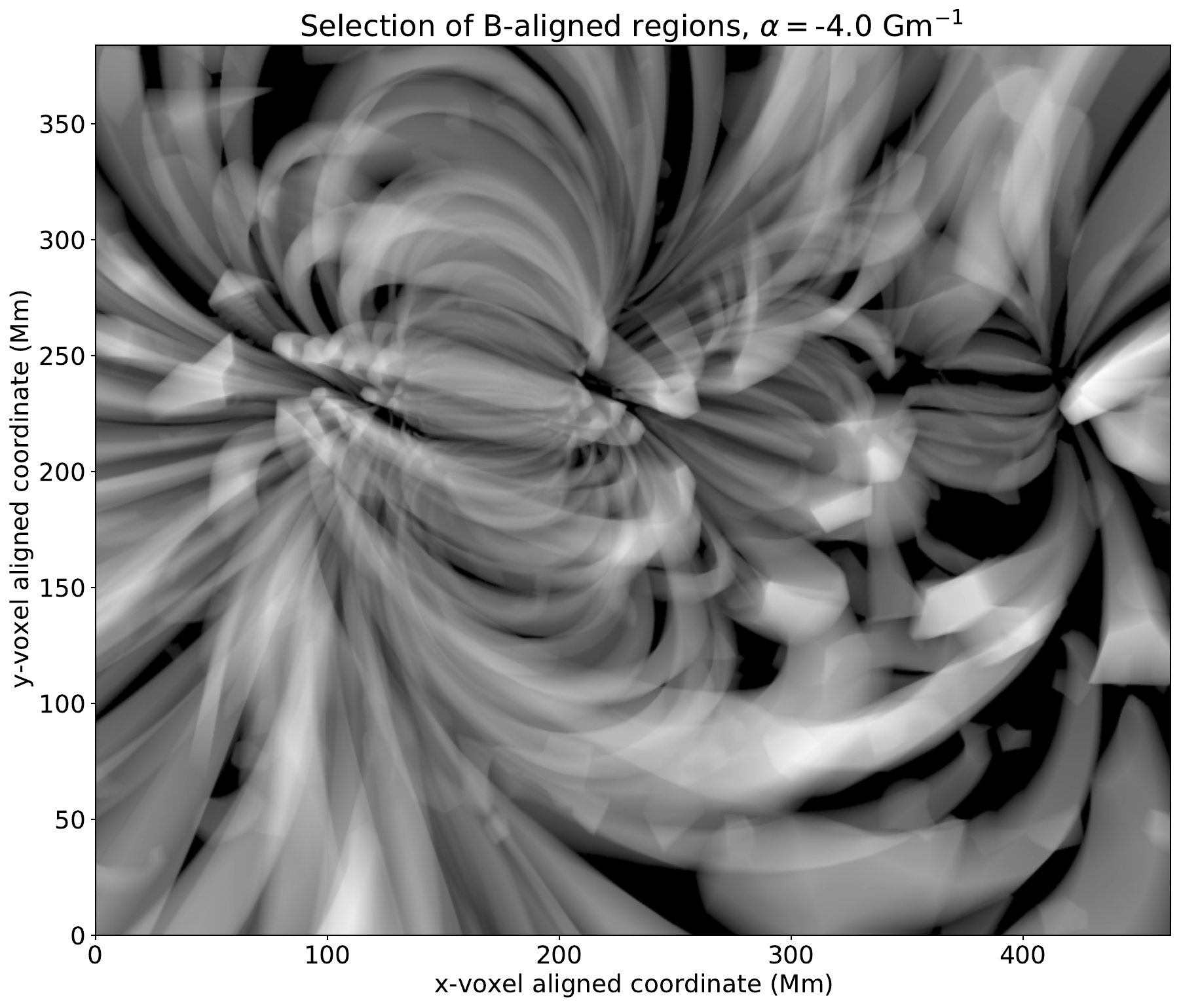}
    \includegraphics[width=0.33\textwidth]{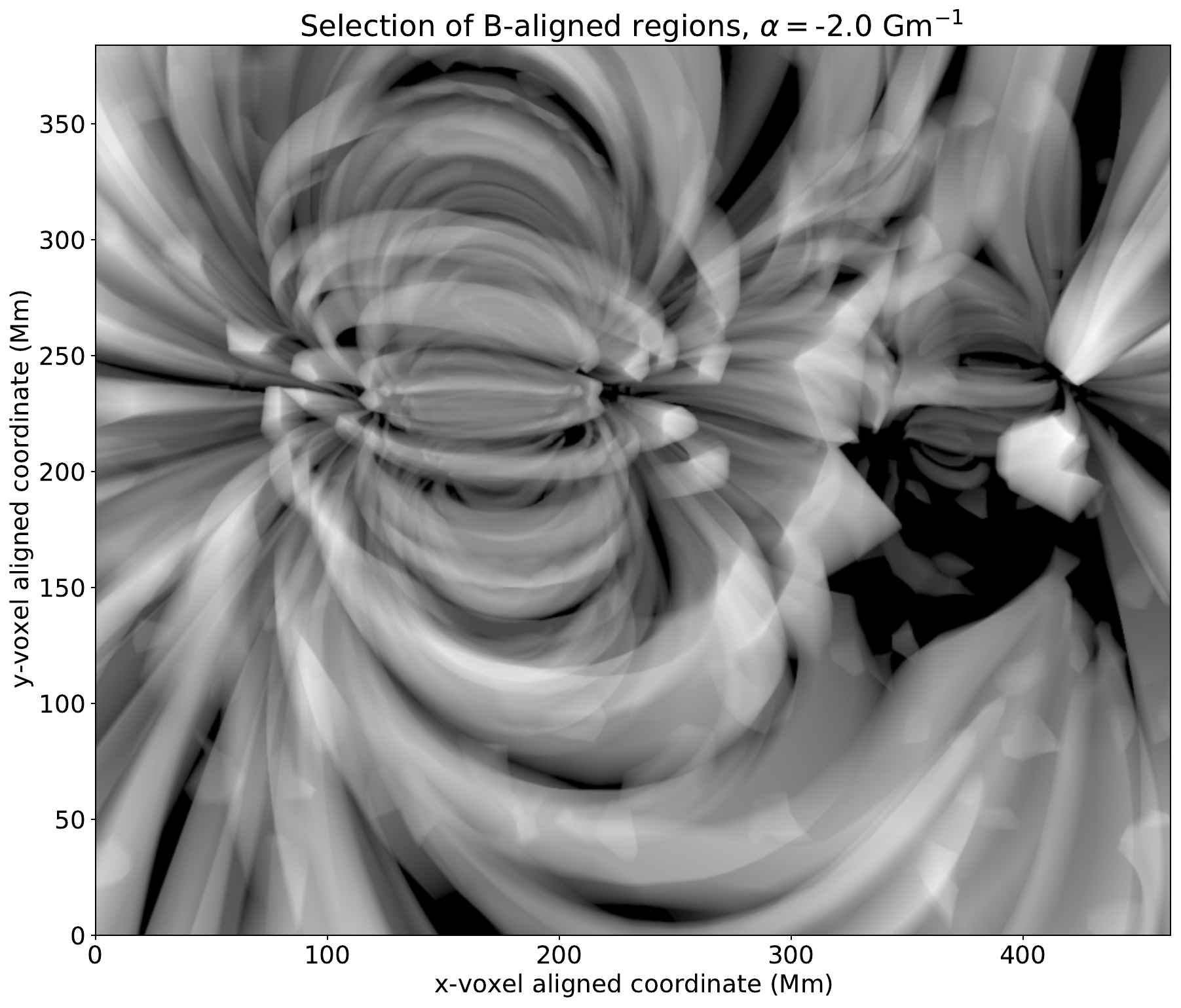}\includegraphics[width=0.33\textwidth]{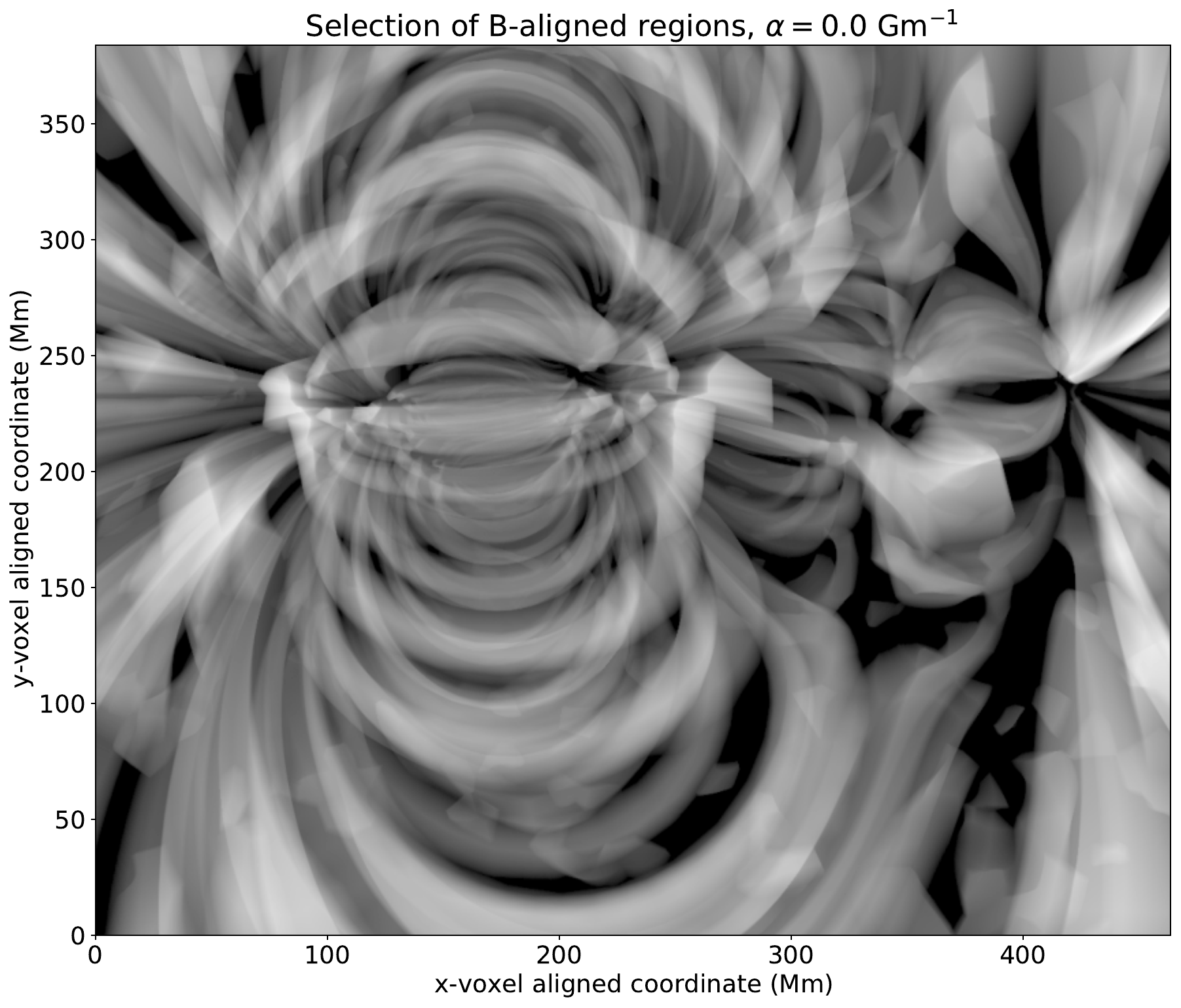}\includegraphics[width=0.33\textwidth]{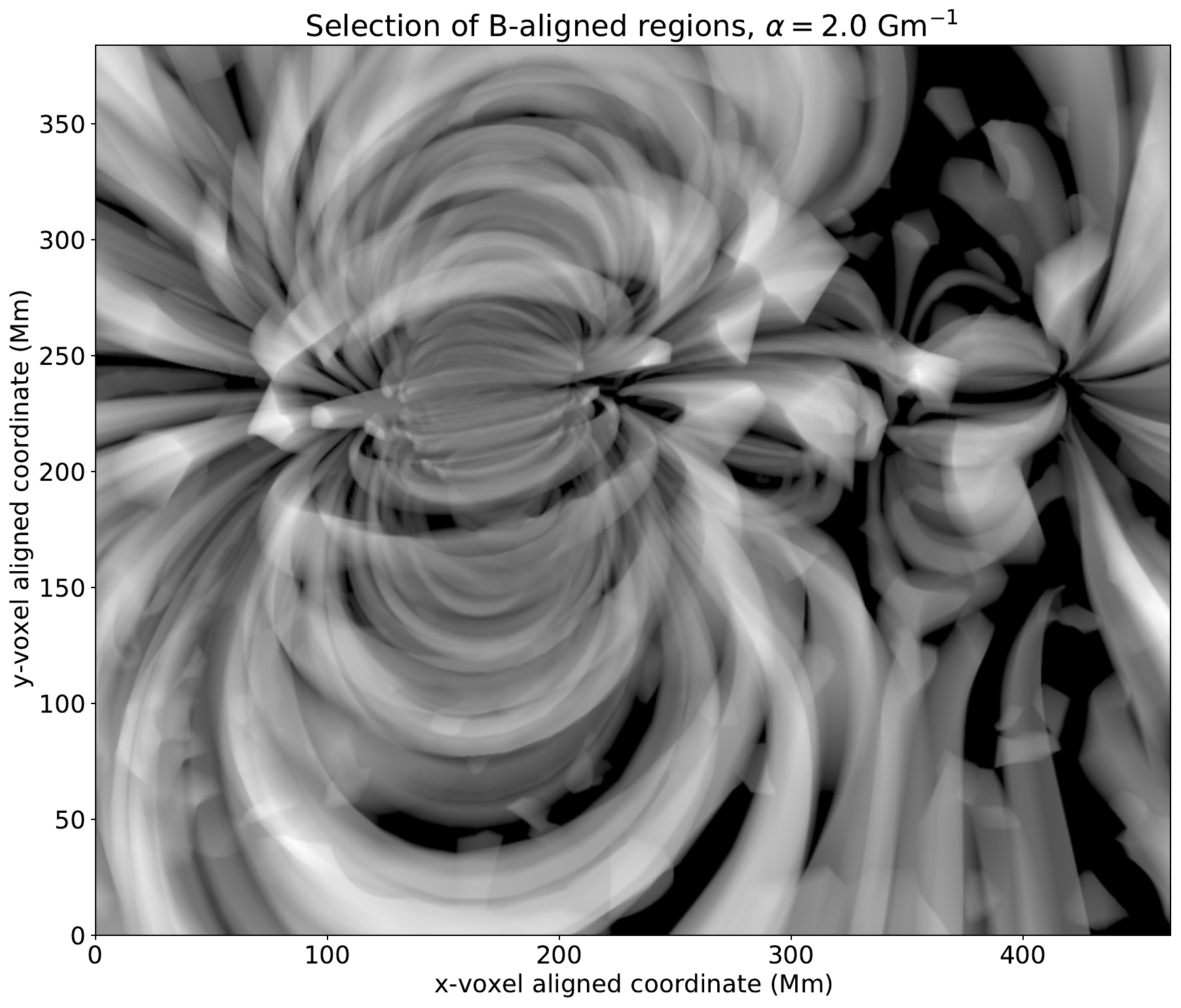}
    \includegraphics[width=0.33\textwidth]{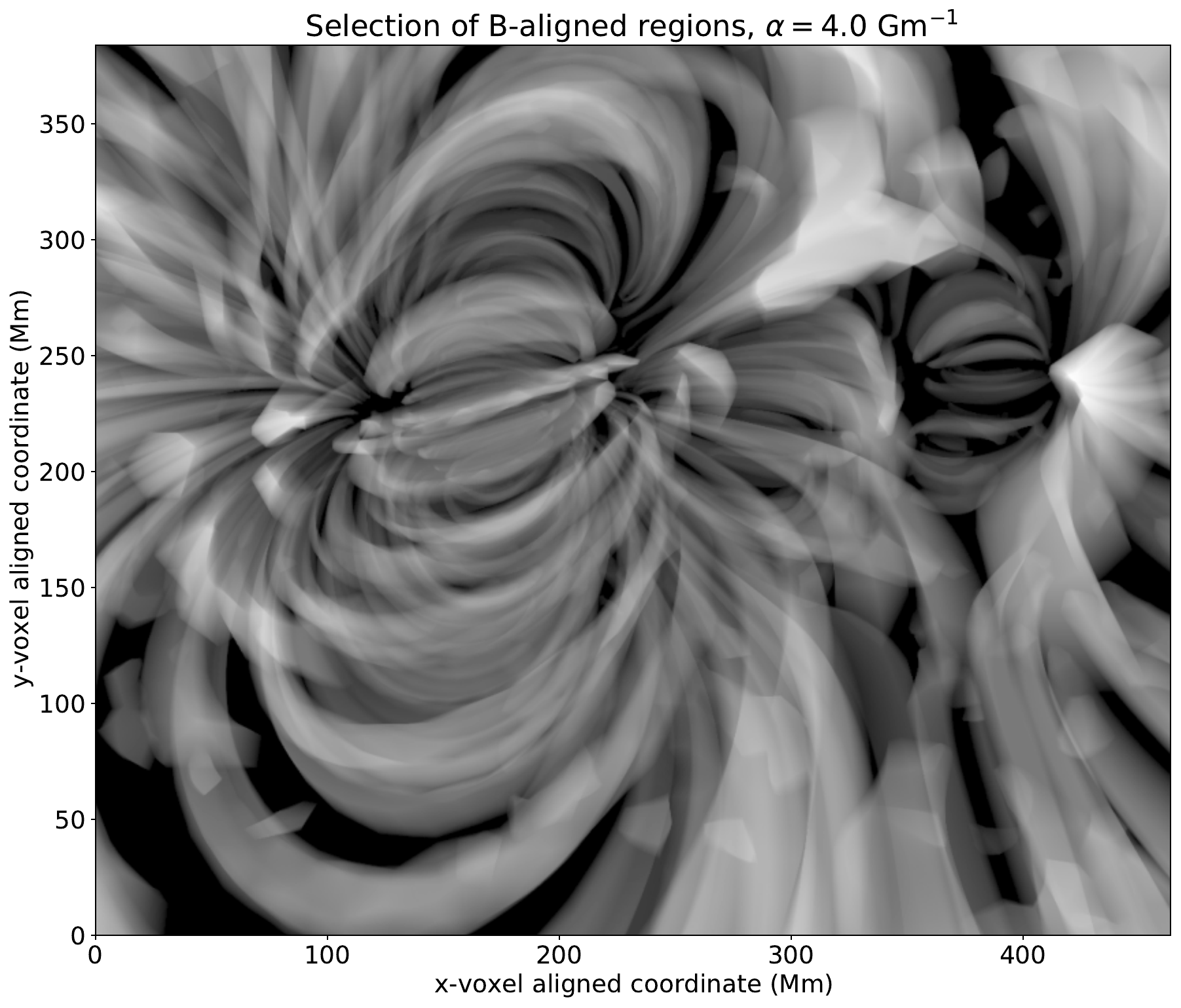}\includegraphics[width=0.33\textwidth]{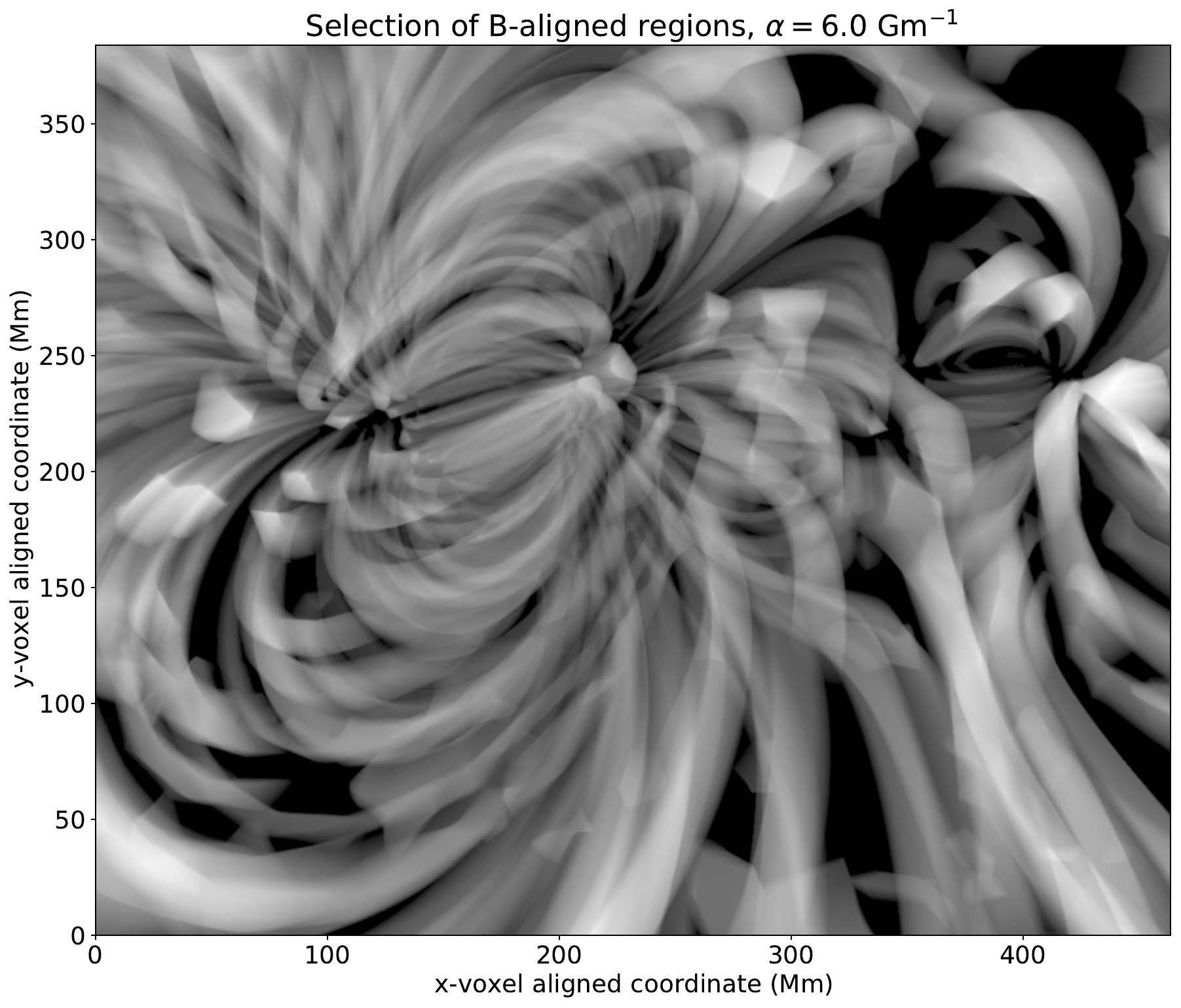}\includegraphics[width=0.33\textwidth]{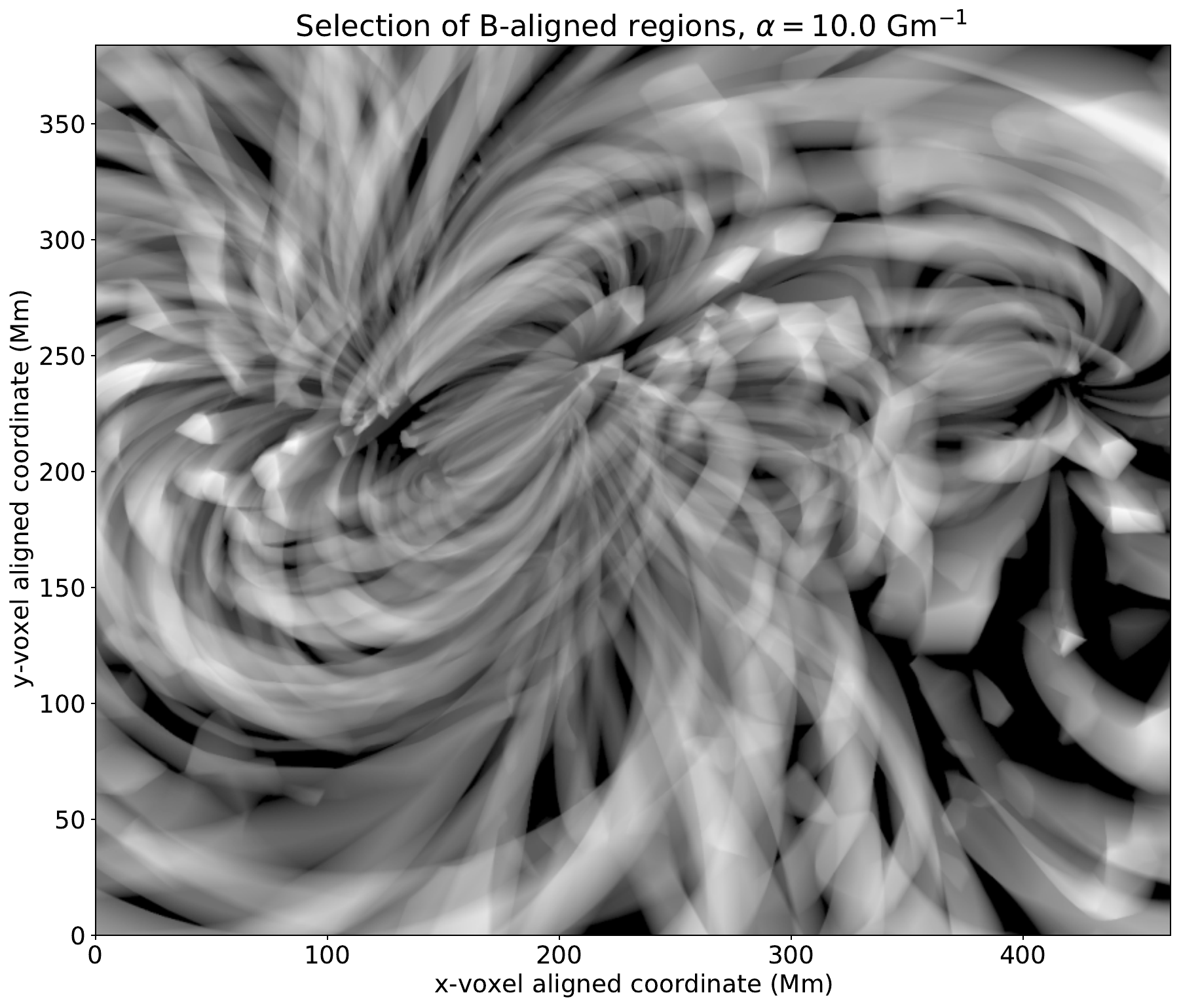}
    \caption{Projection through the voxel volume showing a subset of the $\mathbf{B}$-aligned regions used by CROBAR, with varying values for the $\alpha$ parameter of the linear force-free field. Values start at $-10$ turns per Gm (top left) and increase left to right, top to bottom, to $+10$ turns per Gm (bottom right). $\alpha=0$ is the potential field. The coordinate system is directly overhead of the center of the images, with x- and y-axes aligned with longitude and latitude (the data's `local' coordinate frame.}\label{fig:example_lfff_alpha_loops}
\end{figure}

Part of the process of LFFF extrapolation is finding the best $\alpha$ helicity parameter for the extrapolation. The choice of $\alpha$ determines how much the field lines wrap around each other, and in what sense (clockwise or counterclockwise), where $\alpha=0$ is identical to the potential field case. Figure~\ref{fig:example_lfff_alpha_loops} shows a selection of the field-aligned regions used by CROBAR, from Earth's perspective, with $\alpha$ values ranging from $-10$ to $+10$ turns per Gigameter (the region is $\sim 0.5$~Gm across). The effect of $\alpha$ on the helicity of the field is evident. 

\begin{figure}
    \includegraphics[width=0.33\textwidth]{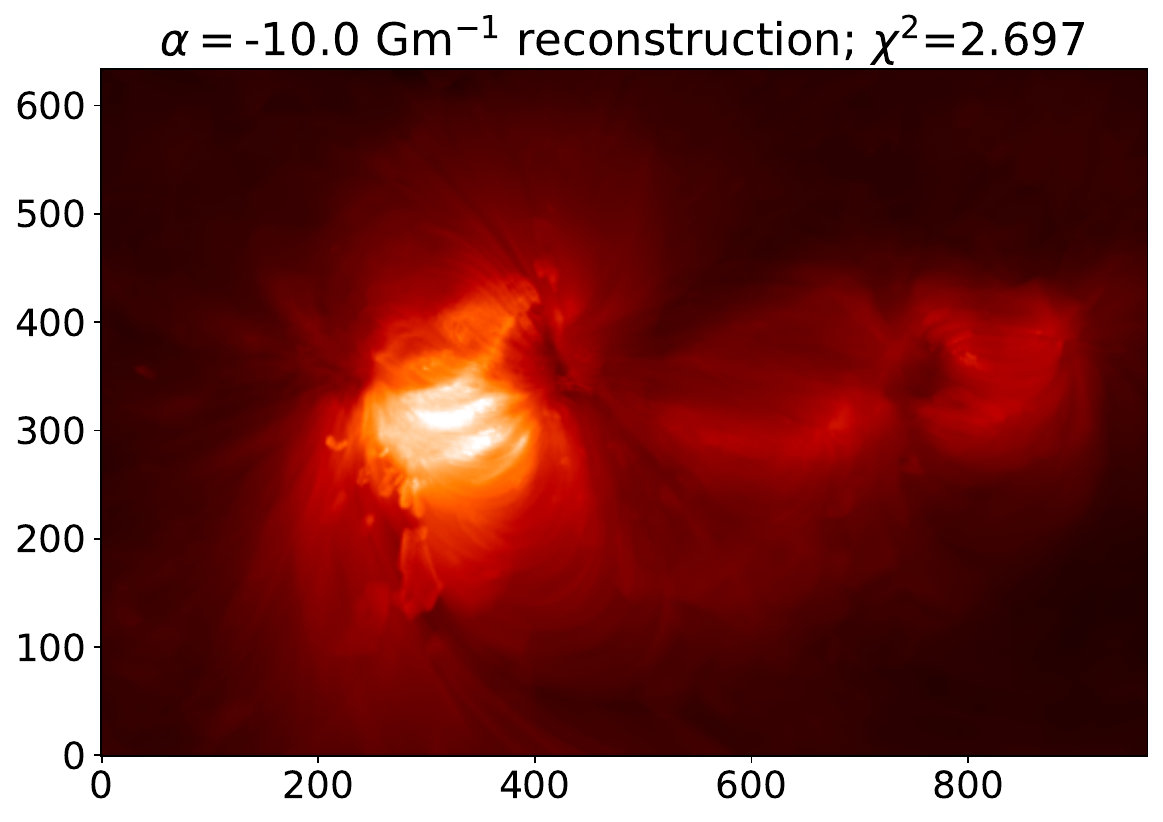}\includegraphics[width=0.33\textwidth]{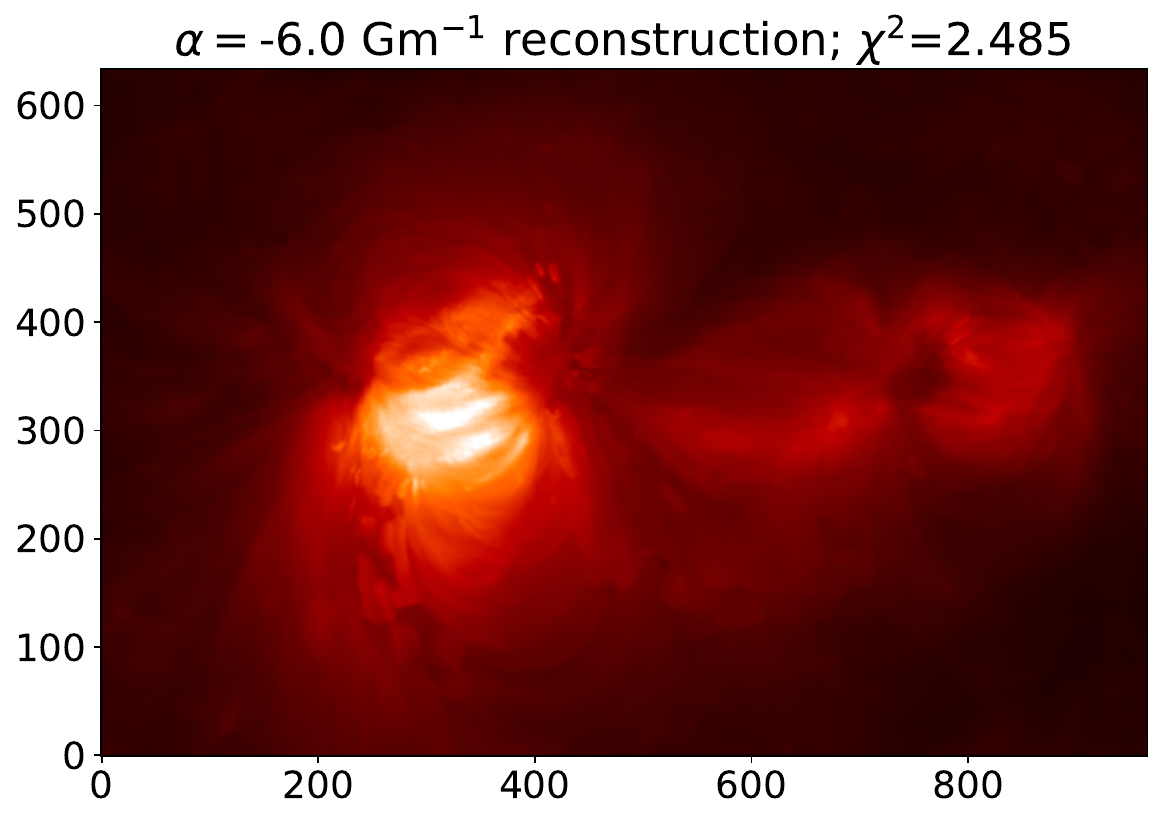}\includegraphics[width=0.33\textwidth]{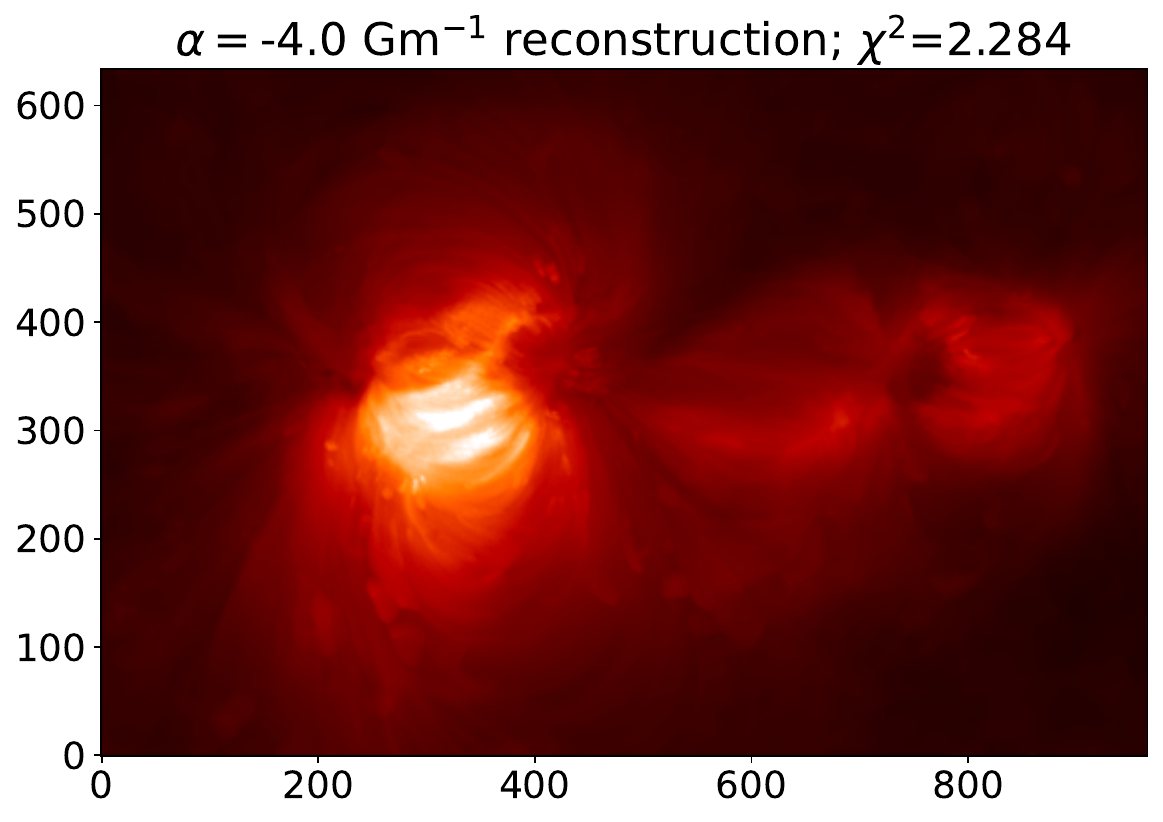}
    \includegraphics[width=0.33\textwidth]{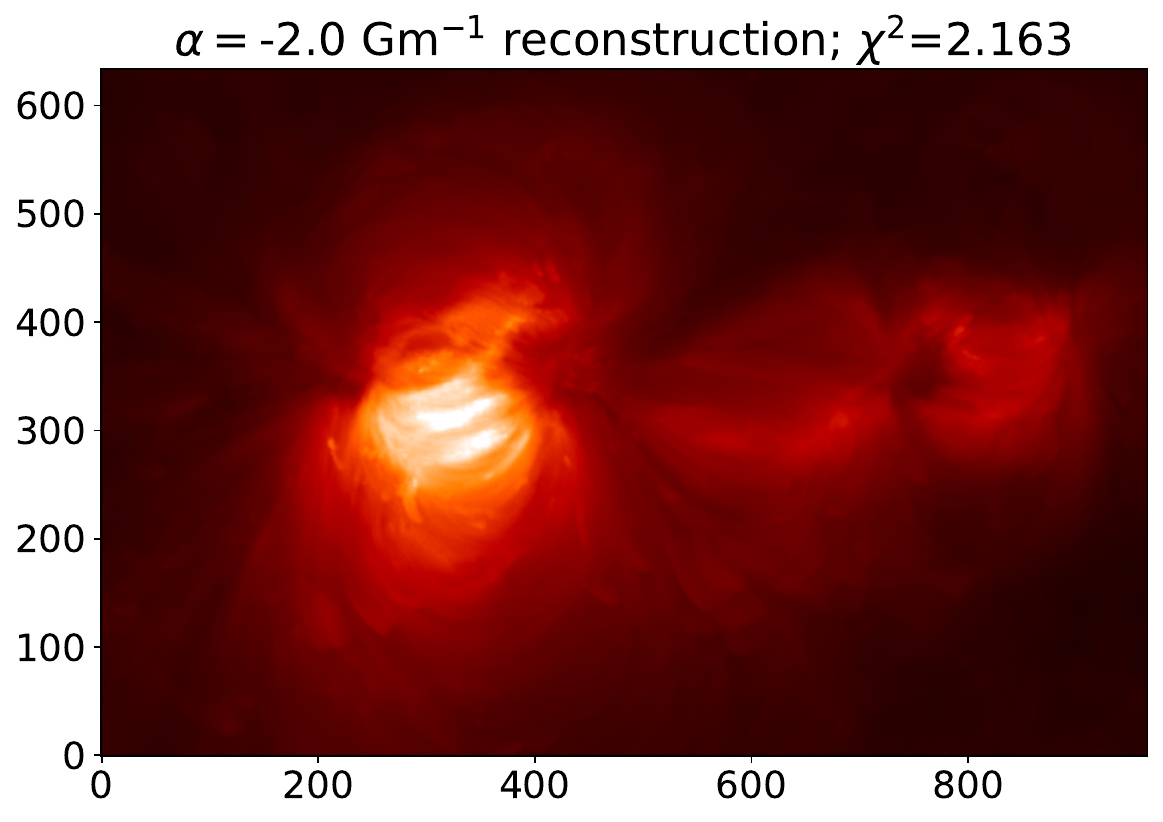}\includegraphics[width=0.33\textwidth]{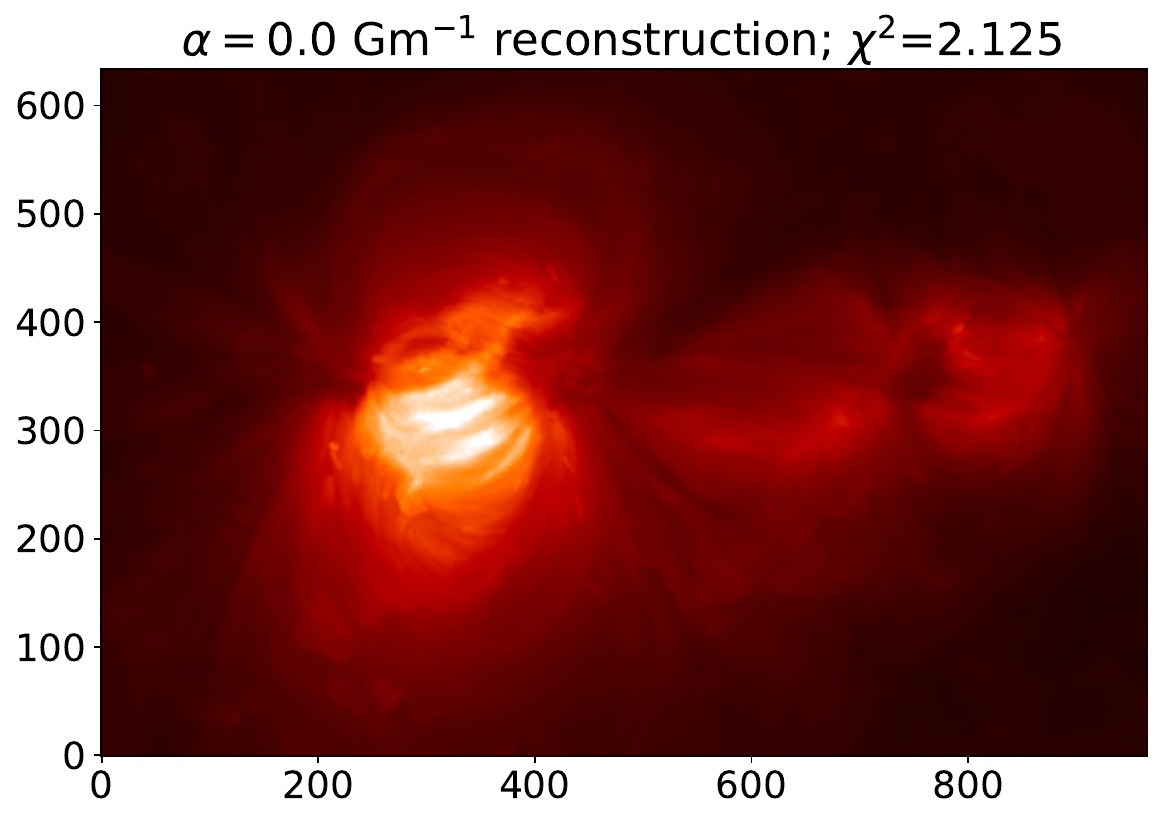}\includegraphics[width=0.33\textwidth]{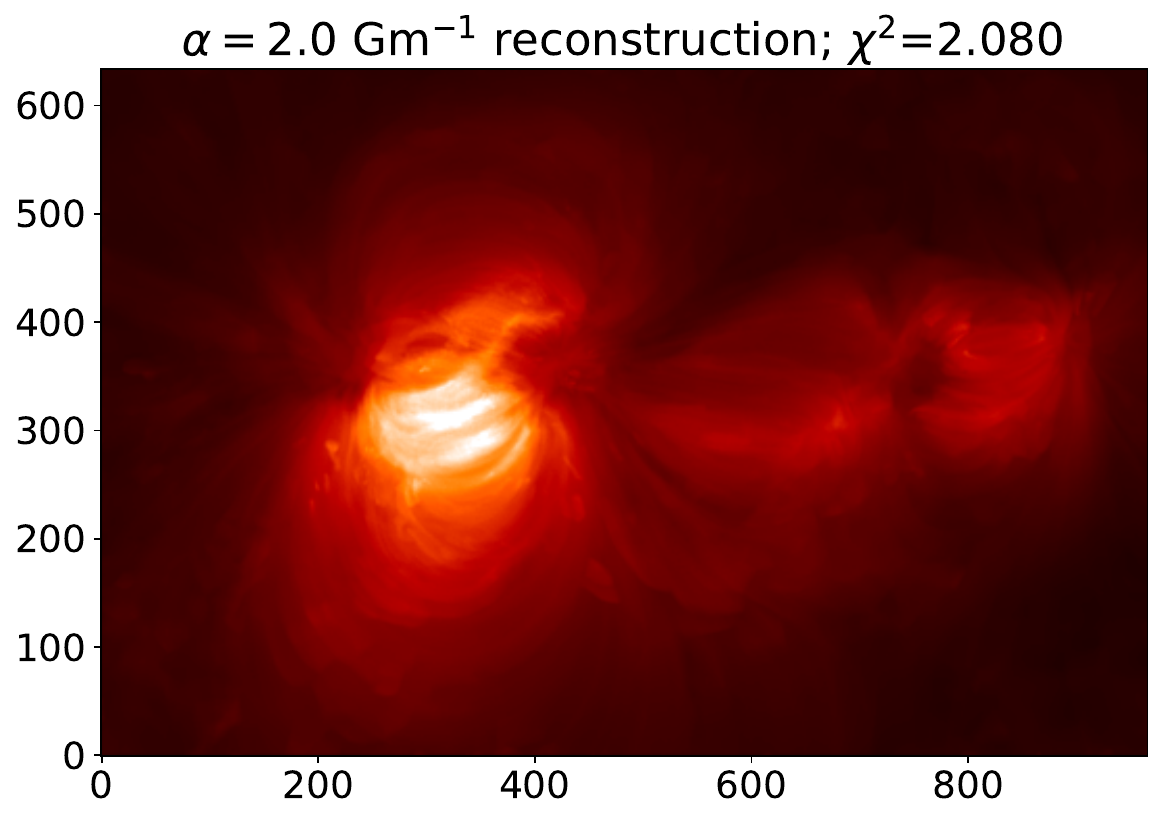}
    \includegraphics[width=0.33\textwidth]{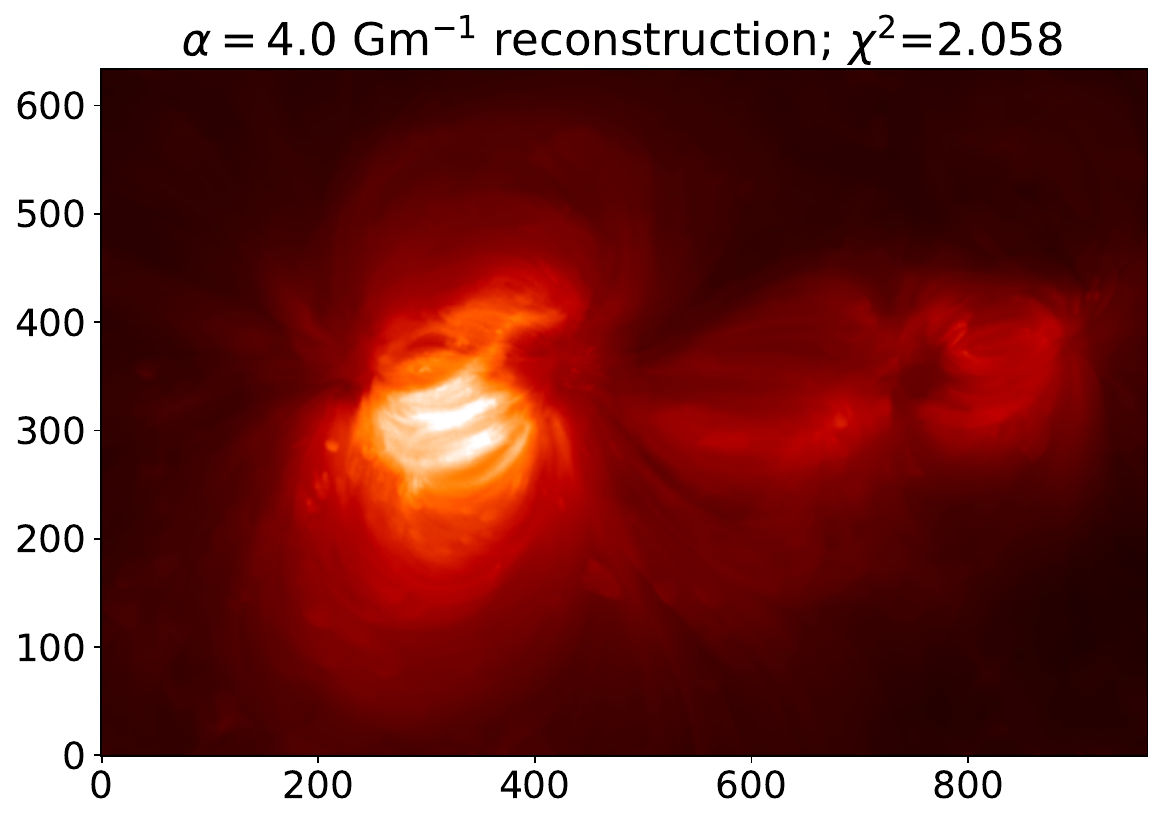}\includegraphics[width=0.33\textwidth]{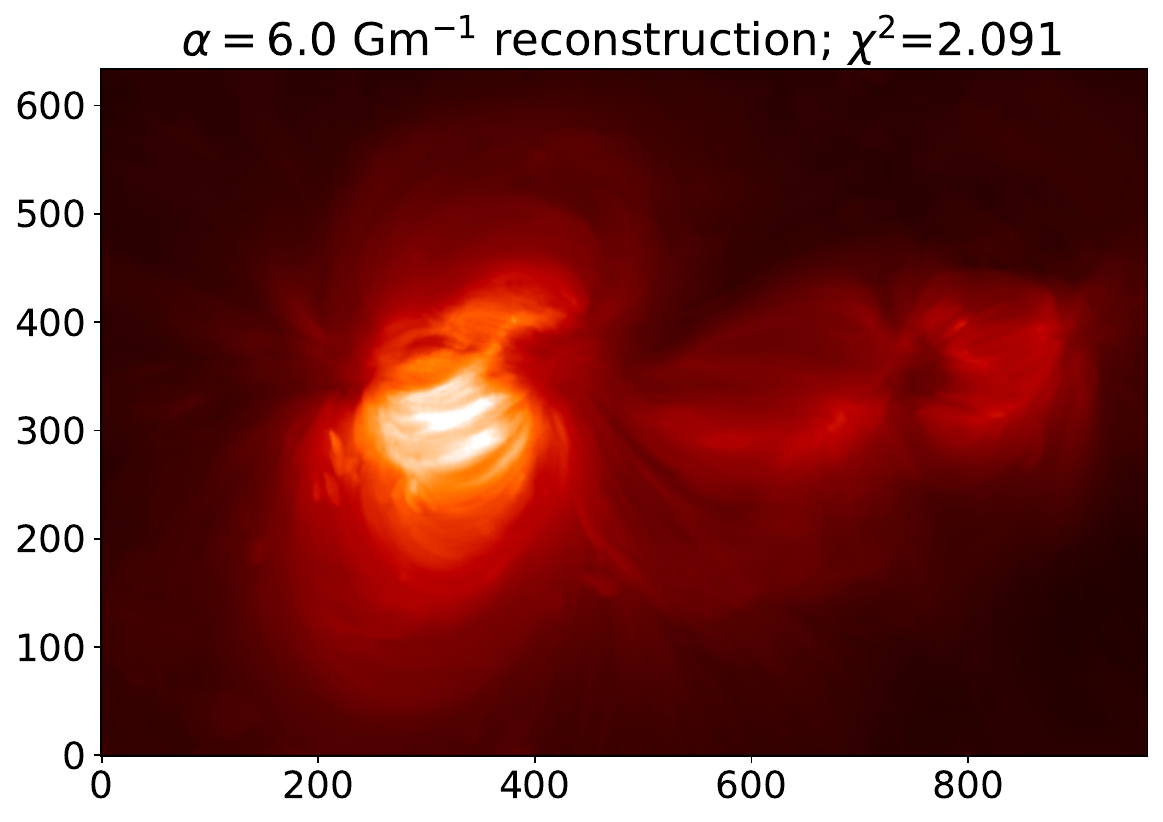}\includegraphics[width=0.33\textwidth]{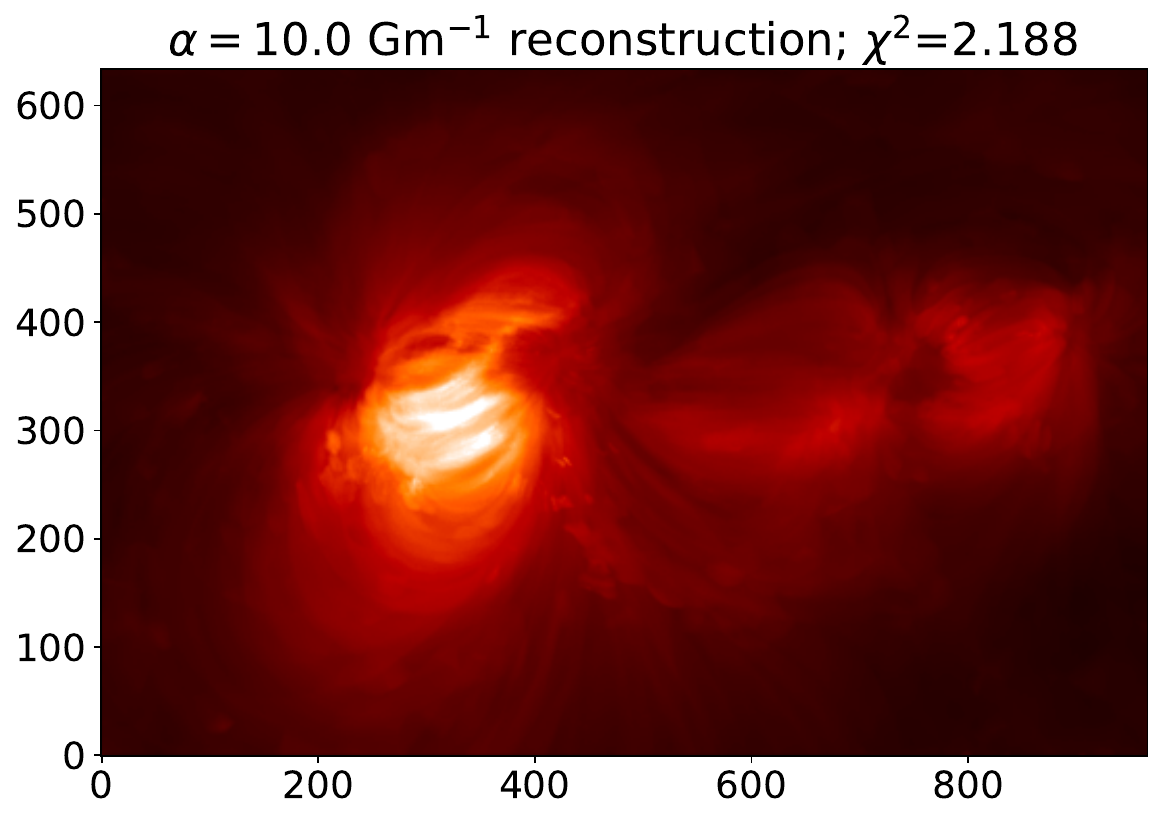}
    \caption{Reconstructions using CROBAR with varying values for the $\alpha$ parameter of the linear force-free field. Values start at -10 turns per Gm (top left) and increase left to right, top to bottom, to +10 turns per Gm (bottom right). $\alpha=0$ is the potential field. The best fit $\alpha$ is roughly 3.5 turns per Gm (see Figure~\ref{fig:best_chi_squared_reconstruction}).}\label{fig:example_lfff_alpha_reconstructions}
\end{figure}

\begin{figure}
    \includegraphics[width=\textwidth]{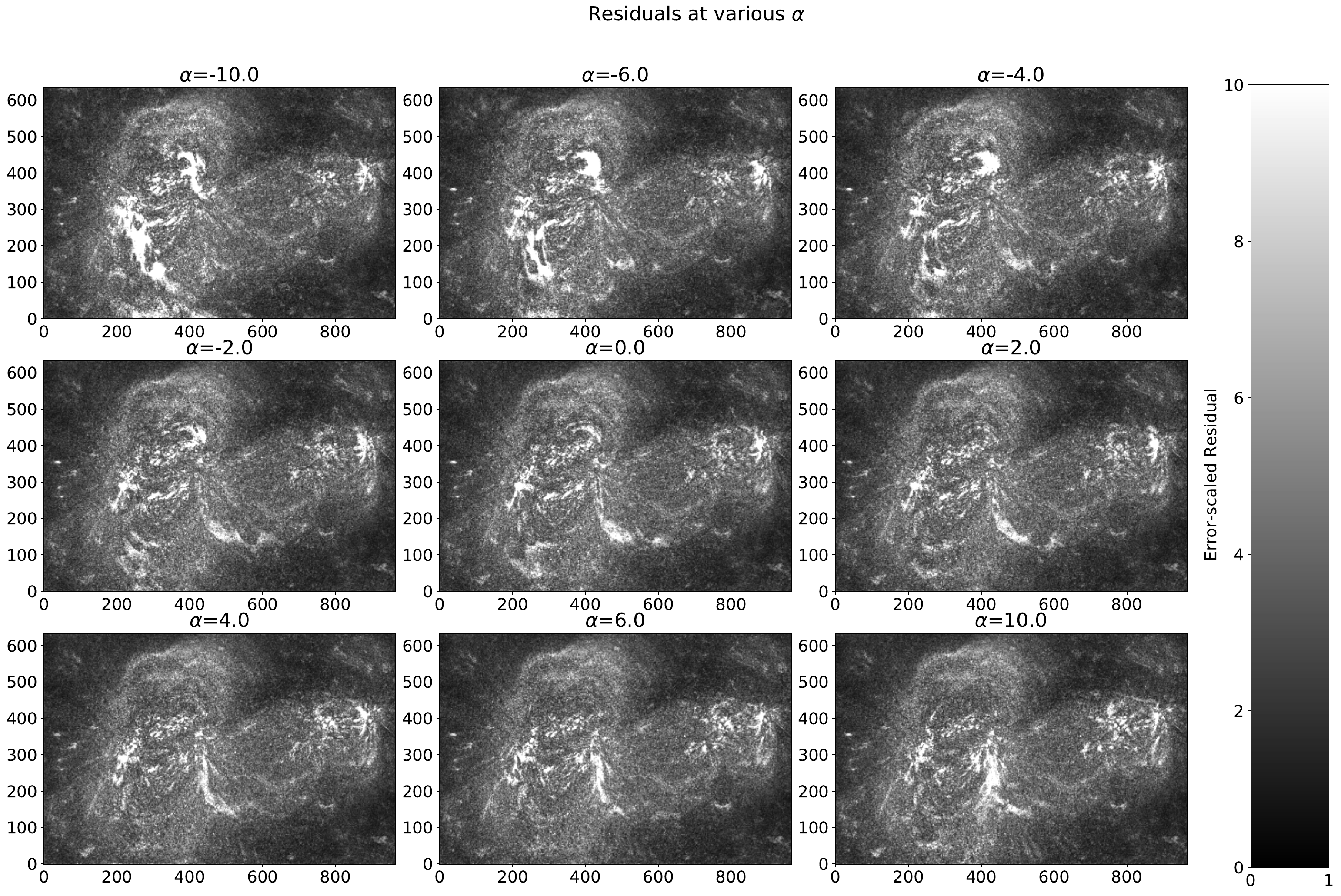}
    \caption{Residuals of the CROBAR fit to the active region, using the $T^2$ power law passband from AIA data, and each of the $\alpha$ values shown in Figure \ref{fig:example_lfff_alpha_loops}.}\label{fig:alpha_residuals}
\end{figure}

One of the signature features of CROBAR is that its reconstructions produce a 3D model of the emission structure that can be compared directly, pixel by pixel, with the EUV images. Therefore the $\chi^2$ residual of the reconstruction can be computed in the standard way by subtracting the modeled image from the original and dividing by the errors in the intensities at each pixel (these errors are dominated by instrumental and photon counting noise). Indeed, it is this $\chi^2$ that drives the reconstruction in the first place, but poorer or better field extrapolations (specifically, choices of $\alpha$) still result in better or worse $\chi^2$ at the end of the reconstruction. This $\chi^2$ therefore allows us to close the loop between the field extrapolation and the optically thin observations. We demonstrate this ability in Figure~\ref{fig:example_lfff_alpha_reconstructions}, which shows CROBAR's reconstructions using the same values of $\alpha$ shown above, along with the $\chi^2$. Figure~\ref{fig:alpha_residuals} show the residuals -- the differences between the reconstruction and the original AIA data, divided by the estimated error, at each pixel -- from which each of these $\chi^2$ were computed. It is evident that the $\alpha=4$ reconstruction works best both in terms of $\chi^2$ and visually. 

\begin{figure}
    \includegraphics[width=0.5\textwidth]{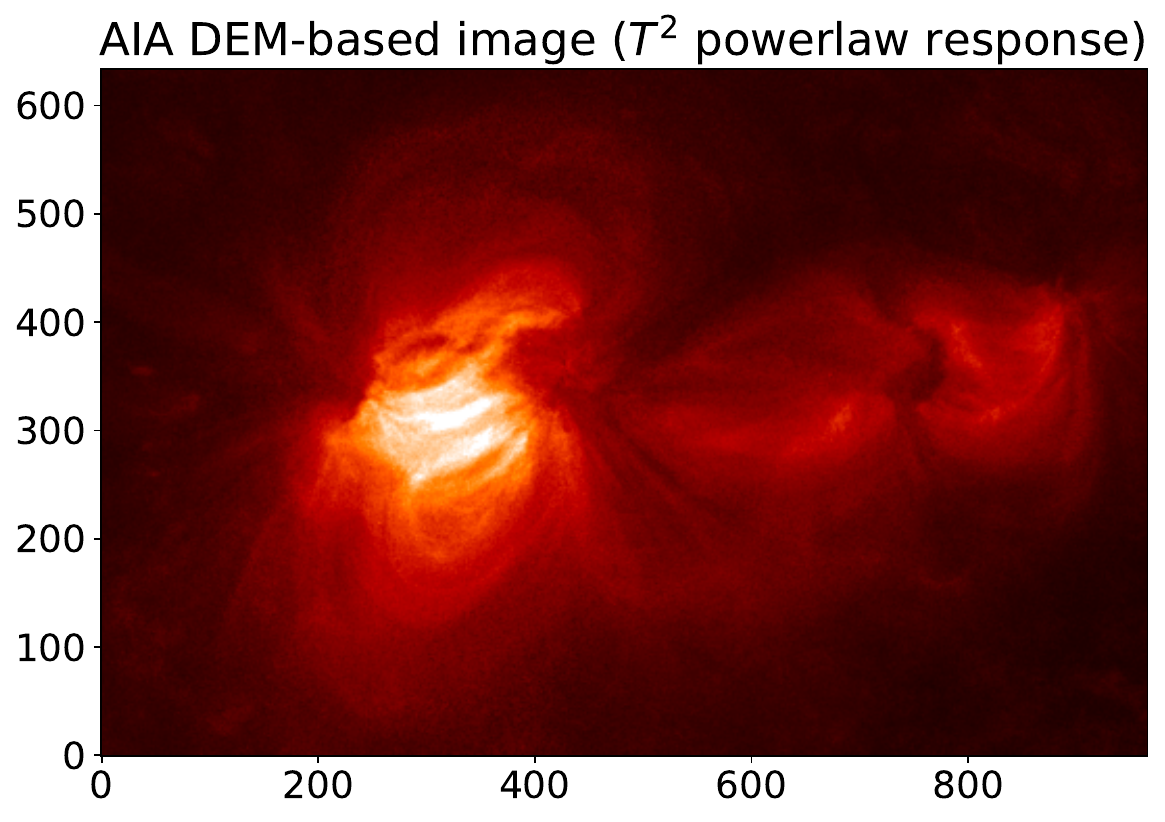}\includegraphics[width=0.5\textwidth]{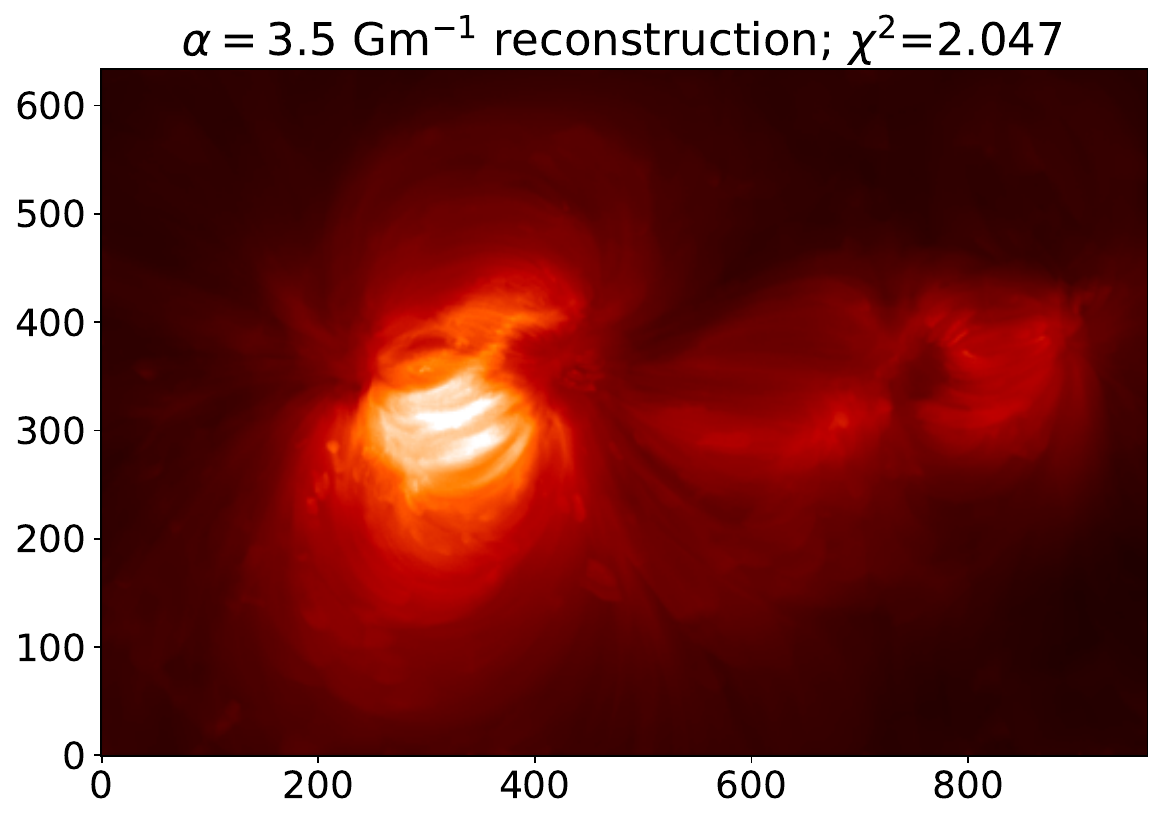}
    \caption{Visual comparison between AIA and CROBAR reconstruction for this region. Original AIA DEM based $T^2$ powerlaw image on left, CROBAR reconstruction of same image with approximate best-fit value of $\alpha$, 3.5 is on right.}\label{fig:best_chi_squared_reconstruction}
\end{figure}

The value of $\alpha$ that produces the best $\chi^2$ can be further refined, either algorithmically (with a Newton-Raphson method, for instance) or by hand, to find the best $\chi^2$ overall. Here we have performed the search by hand, and find that the best fit value is roughly 3.5; Figure~\ref{fig:best_chi_squared_reconstruction} shows the corresponding reconstruction alongside the AIA image once again

\begin{figure}
    \includegraphics[width=\textwidth]{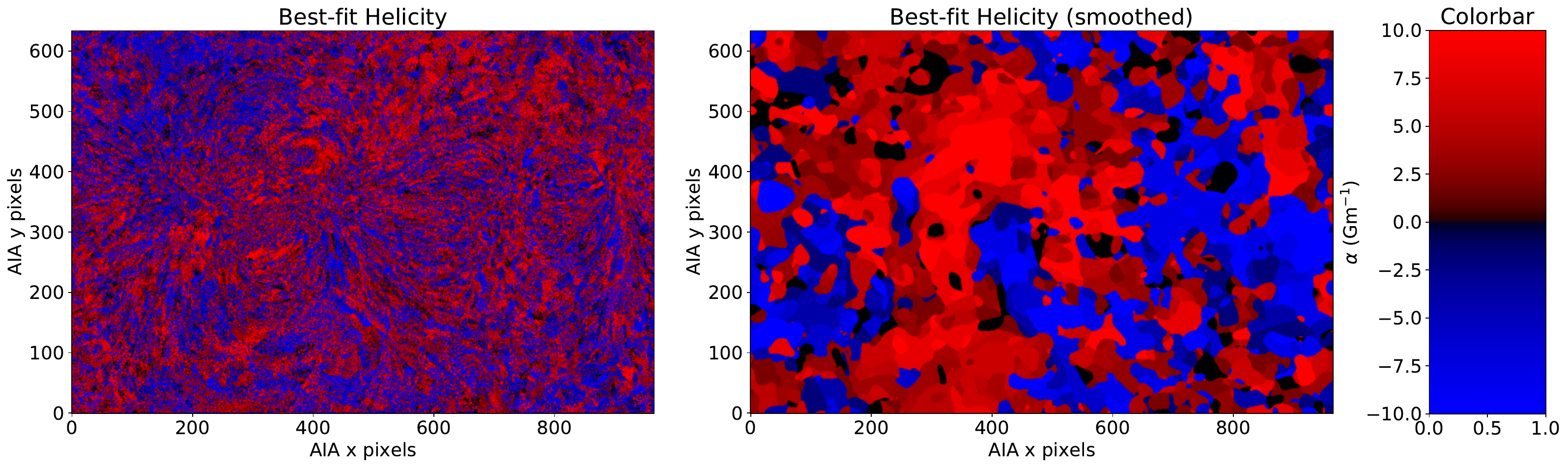}
    \includegraphics[width=\textwidth]{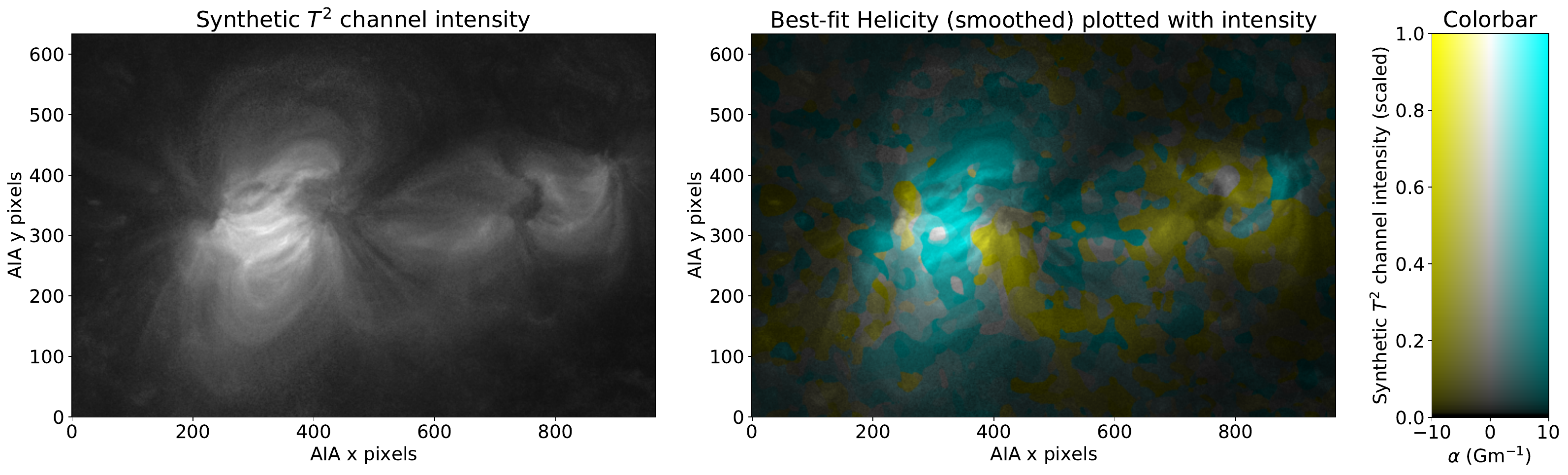}
    \caption{Helicity of the region estimated from CROBAR reconstruction. Top left: helicity map without smoothing, demonstrating noise in the result. Top right: smoothed helicity. In the top row, positive helicity is shown in red, negative helicity is shown in blue. Lower rows show the smoothed helicity again with the intensity for comparison. Lower left: intensity used to make the reconsutrction. Lower right: Helicity (negative in yellow, positive in cyan) with intensity (brightness) shown on the same figure.}\label{fig:alpha_map}
\end{figure}

The `linear' assumption in the LFFF is that the same constant of proportionality applies across the entire extrapolation region, but in reality the Sun has no obligation to follow this assumption. As a result, the $\alpha$ that best fits the data can vary from one part of the image to the next. We can capitalize on this by producing an image of which $\alpha$ produces the lowest $\chi^2$ at each pixel.  The result is a map of which $\alpha$ best fits the data across the entire image, providing a straightforward to identify regions of high helicity, and therefore high amounts of free magnetic energy (in future work we will incorporate nonlinear Force Free Field, or nLFFF, extrapolations into CROBAR, which are the fully self-consistent realization of spatially varying $\alpha$). However, there is a significant amount of noise in this result, both due to measurement error from the AIA observations and due to subsections of loops happening to be a better fit at particular places rather than an entire field region being a better fit. To counteract this, we have smoothed the residuals, and used that as the basis of a spatially varying $\alpha$ map -- in effect, showing which $\alpha$ best fits the observations over the smoothing region. This is shown in Figure~\ref{fig:alpha_map}, using a Gaussian smoothing kernel of $\sigma = 10$ AIA pixels. 

\begin{figure}
    \includegraphics[width=\textwidth]{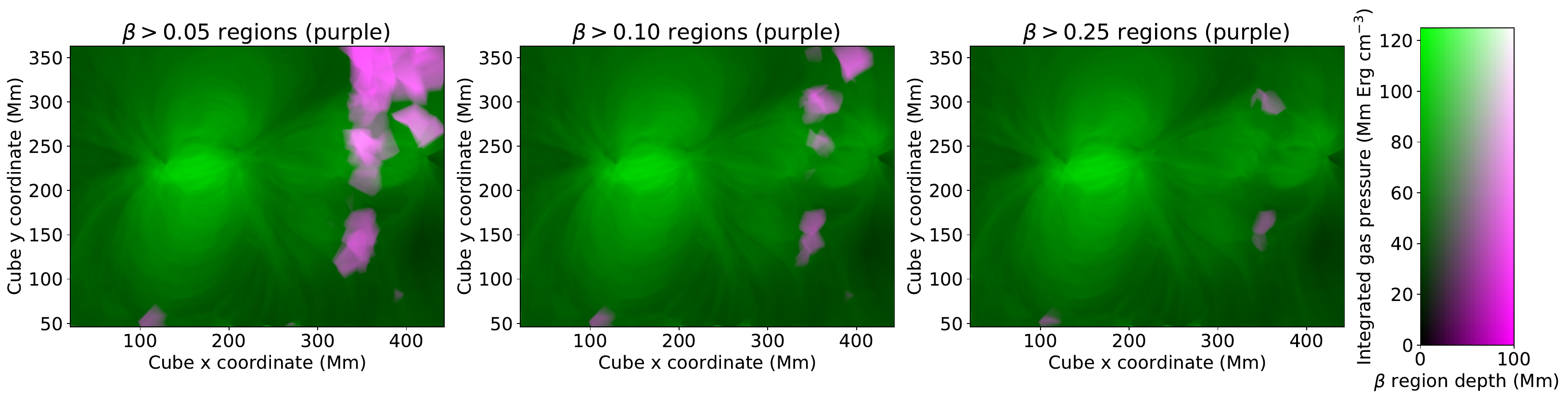}
    \includegraphics[width=\textwidth]{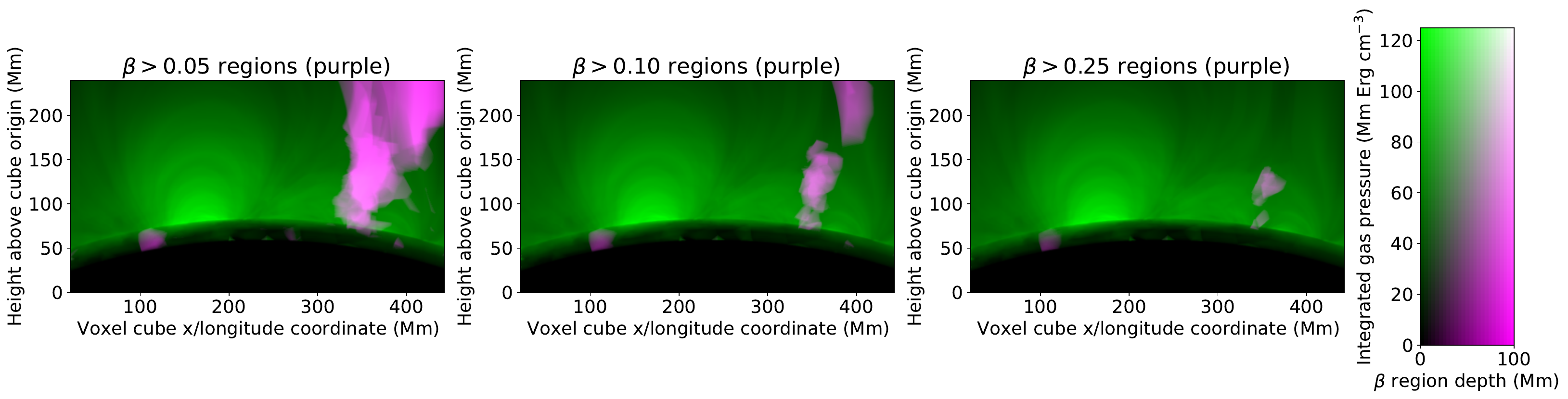}
    \includegraphics[width=\textwidth]{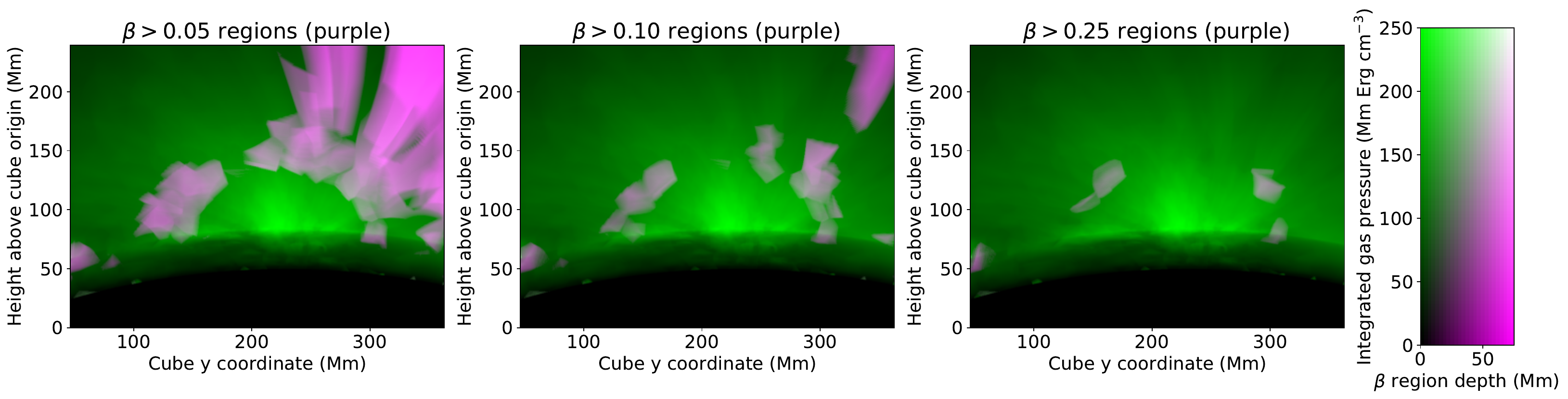}
    \caption{Regions of (relatively) high plasma $\beta$ (in purple) shown plotted over the volume-integrated plasma pressure (in green). Top row shows the z (vertical) axis projected view, second row shows the y (latitude-aligned, roughly polar) axis projected view, third row shows the x (longitude-aligned) axis projected view. Each column shows regions falling above a threshold of increasingly high plasma $\beta$: 0.05 on left, 0.1 at center, 0.25 on right.}\label{fig:field_energetics}
\end{figure}

Furthermore, the reconstruction of the power-law index 2 synthetic channel is essentially a reconstruction of the square of the pressure: With that temperature response, each volume element contributes an intensity proportional to
\begin{equation}
    R(T)n^2 dl = T^2 n^2 dl \propto \frac{P^2}{n^2} n^2 dl = P^2 dl
\end{equation}
to the line of sight integral forming the synthetic channel image, where we have used the ideal gas law ($P \propto nT$) to convert temperature into the ratio of pressure and density. Therefore, we can estimate the plasma $\beta$ -- that is, the ratio of gas dynamic pressure to magnetic pressure -- for the region via a straightforward ratio of this reconstruction and the magnetic pressure we compute from the field extrapolation. $\beta$ is a critical tracer of locations where magnetic energy release can initiate (via ideal MHD instabilities or reconnection), since it measures the relative importance of the magnetic field and gas dynamic physics in determining the behavior of the plasma. In much of the corona, $\beta$ is low (${\ll}1$) and therefore the magnetic field drives the processes. Where $\beta$ is near unity, on the other hand, gas dynamic behaviors can drive the processes. In particular, the gas dynamic pressure can drive flows that can initiate magnetic reconnection, through which stored magnetic energy can be released in the plasma. CROBAR allows us, for the first time, to visualize plasma $\beta$ in three dimensions, as shown in Figure~\ref{fig:field_energetics} along with the plasma pressure.

Generally, the plasma $\beta$ in this reconstructed region is low, with the exception of the smaller region to the right of the image. This appears to have a pair of magnetic nulls (locations with zero magnetic field) above and below it connected by a separator (or Quasi-Separtrix Layer) and to be associated with a pseudo-streamer when viewed from the roughly polar perspective. For this region, this is a perspective not accessible to any current instrumentation, and accessible only with CROBAR. It is not obvious that the region would have this structure based on visual inspection of the original AIA data, but becomes clear thanks to CROBAR's ability to combine the magnetic and optically thin information into a true three-dimensional picture that also captures the underlying physics of the coronal plasma.

Finally, we can also map the free magnetic energy in the region, which is the difference between the magnetic energy in the force-free field(s) and in the potential field. The potential field is the minimum energy state in the corona given the observed photospheric field configuration, thus the difference in magnetic energy ($B^2/2$) between other field configurations and the potential configuration is the energy free to do work on the field's environs, given that the observed photospheric field configuration remains fixed. (Naturally, it is possible that more energy could be liberated if a coronal event allows further rearrangement of the field in the photosphere, but such a `tail wagging the dog' event is not likely to result in a major rearrangement of the photospheric field and the energy so liberated will largely remain in the photosphere).

\begin{figure}
    \includegraphics[width=\textwidth]{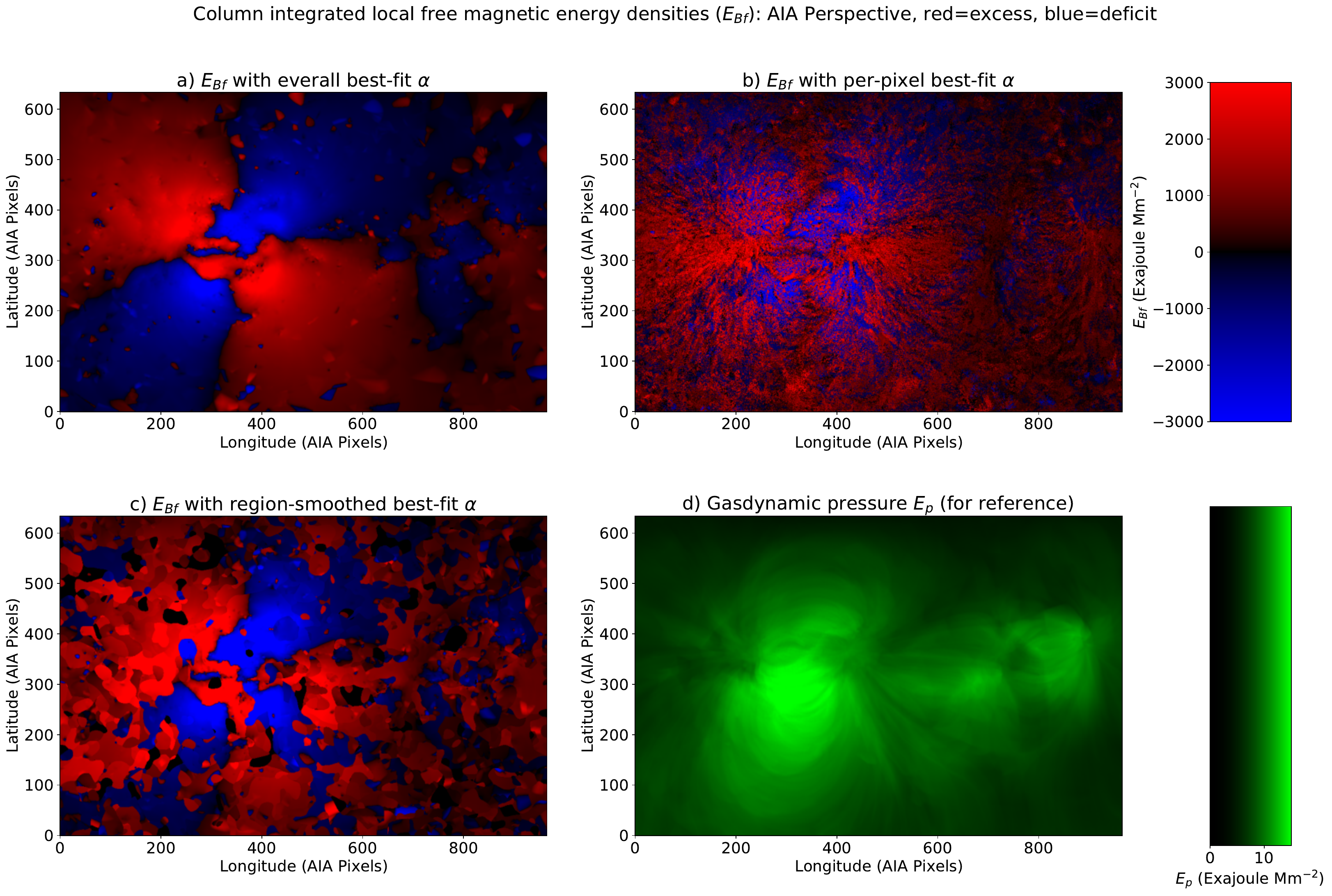}
    \caption{Free magnetic energy in the CROBAR reconstructions, as seen from AIA's perspective. In this paper, a surplus (local) free magnetic energy is shown in red, while a deficit is shown in blue. (a) Using the overall best-fit $\alpha$. (b) Using the per-pixel best $\alpha$. (c) Using the best $\alpha$ over 10-pixel-radius Gaussian-smoothed regions. (d) The gas dynamic pressure/energy density is shown for reference.}\label{fig:emag_composites}
\end{figure}

\begin{figure}
    \includegraphics[width=\textwidth]{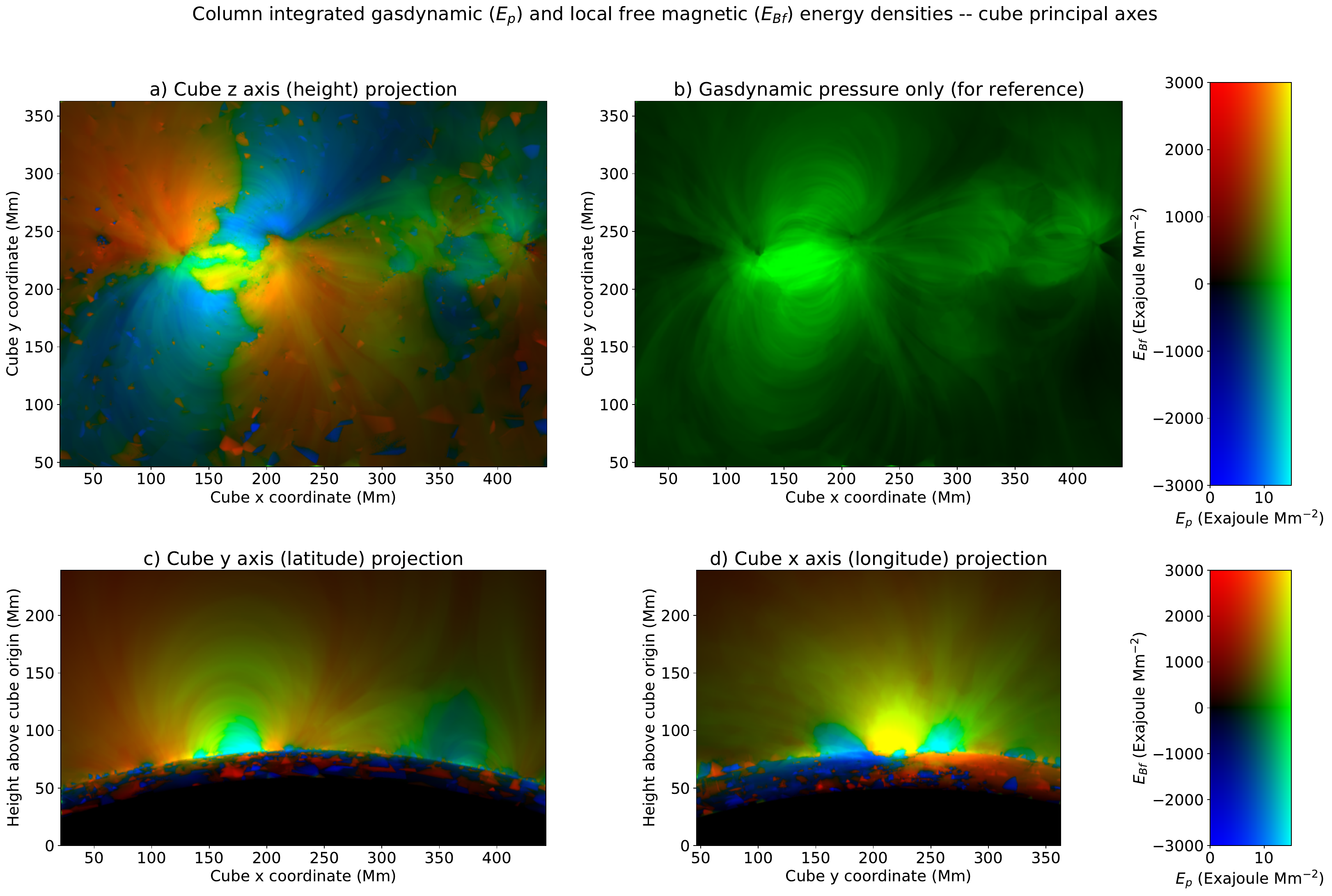}
    \caption{Free magnetic energy in the CROBAR reconstructions, viewed along the three principal axes of the reconstruction cube. All are using the overall best for value of $\alpha$, 3.5 Gm$^{-1}$. (a) overhead/height projection. (b) Same perspective with plasma pressure only, for reference. (c) y axis (latitude) projection. (d) x axis (longitude) projection. A surplus in the (local) free magnetic energy is shown in the red channel, while a deficit is shown in the blue channel. The plasma pressure is shown in the green channel. In this color scheme, a region with a large amount of excess magnetic energy but low plasma energization will appear red, a large energy deficit and low plasma energization will appear blue, large excess magnetic energy and high plasma energization will appear yellow, and large energy deficit with high plasma energization will appear cyan. }\label{fig:emag_perspectives}
\end{figure}

We show this free magnetic energy in Figures~\ref{fig:emag_composites} and \ref{fig:emag_perspectives}, first from the AIA perspective with the single, per-pixel, and per-pixel smoothed best-fit alphas shown in Figure \ref{fig:alpha_map} (Figure~\ref{fig:emag_composites}) and then with the multiple views (along the three principal axes of the reconstruction cube) of the single overall best-fit $\alpha$ in Figure~\ref{fig:emag_perspectives}. In this second figure, the free energy is shown along with the plasma (gas dynamic pressure) energy for reference: it is clear from the figure that the magnetic energy reservoir is much larger than the plasma energy, and that the magnetic energy tends to be concentrated near the footpoints while the plasma energy is distributed more evenly along the field lines.  

It is important to note that the free magnetic energy is a non-local phenomenon, so {\em locally} the non-potential energy can be lower than the potential (i.e., {\em locally} the free magnetic energy can be negative); the \textit{global} free magnetic energy in all cases is positive. We found that the free magnetic energy for the overall best-fit $\alpha$ (3.5 Gm$^{-1}$) was $9\times 10^{31}$, while it was $1.7\times 10^{32}$ ergs if we took the best $\alpha$ pixel by pixel, and $1.1\times 10^{32}$ ergs in the Gaussian smoothed case. The plasma (gas dynamic pressure) energy, on the other hand, was found to be $2.8\times 10^{30}$ ergs. The free magnetic energy reservoir is therefore larger by a factor of $\sim$30 than the plasma energy. In the energy figures in this paper, we use units of Exajoules (EJ) as they provide the closest order of magnitude to represent these energies. One EJ is $10^{25}$ erg, the energy of a large nanoflare or a small microflare.

CROBAR therefore provides a new way to diagnose how energy is stored in the corona, and can also be used to observe how energy is moved around and how it's released. It is a brand new visualization of how these properties are correlated with each other, allowing them to be placed within the coronal volume in three dimensions, in context with one another.

\section{Multiple Perspectives}
\label{sec:multiple_perspectives}
We now turn our attention to the question of multiple perspectives. With the matrix form used by CROBAR, it is straightforward to incorporate multiple simultaneous perspectives in the solution, as we discuss below. But first we will touch on the data products available for demonstrating this capability.

\subsection{Overview of SDO and STEREO}
\label{sec:SDO_STEREO}
Until recently, the only source of observations of optically thin plasma from off of the Earth-Sun line has been the twin-spacecraft STEREO mission. Although Solar Orbiter data are now available as well, the higher-temperature STEREO passbands are more similar to the power-law-index 2 temperature response functions that are ideal for CROBAR. The EUVI 284~\AA\ passband in particular is quite similar to AIA 335~\AA\ and to the power-law 2 function, as is shown in Figure~\ref{fig:AR_example_image}. The angles of the STEREO spacecraft were increasingly far from those of AIA for much of AIA's observing period (beginning in 2010), and STEREO B was also lost in 2014. The best time for simultaneous observations was therefore at the very beginning of the AIA mission in 2010. We have identified the active region, already shown, which was observed on July 25, 2010, when the two STEREO spacecraft were about $\pm 70^\circ$ from the Earth-Sun line, as the best period for a multiple vantage point test. The presence of both spacecraft allows an independent verification of a two-vantage-point reconstruction: EUVI-A \& AIA can be validated with EUVI-B, and EUVI-B \& AIA can be validated with EUVI-A.

STEREO does not have enough EUV channels to perform a reliable DEM reconstruction from its data alone, so we cannot re-use the $T^2$ synthetic passband previously shown. Instead we use the EUVI 284\,\AA\ passband. Because the 284\,\AA\ passband of STEREO is not identical to the 335\,\AA\ AIA one, however, we do use the DEMs from AIA to synthesize a pseudo-AIA 284\,\AA\ passband. All three perspectives of 284\,\AA, including the AIA one synthesized from DEMs, are shown in the figures next to the respective reconstruction results: Figure \ref{fig:stereoa_reconstructions} for STEREO A, Figure \ref{fig:stereob_reconstructions} for STEREO B, and Figure \ref{fig:aia_stereo_reconstruction} for AIA. We found that a reconstruction simply using AIA 335\,\AA\ directly was not quite as good due to the passband mismatch with EUVI, even though the AIA 335\,\AA\ passband is otherwise a slightly better match for the $T^2$ power law temperature response (see Figure~\ref{fig:AR_example_image}).

\subsection{How Perspectives are Combined}
\label{sec:perspective_combining}
Once the forward matrix mapping from the coefficients of the $\mathbf{B}$-aligned regions to the individual pixels is set up, it is a straightforward matter of applying the linear least squares (with nonlinear mapping for positivity) formalism to obtain the best-fit solution. This is a standard matrix inversion exercise, and all of the physical coordinate system and coordinate transform aspects of the problem have been encapsulated in the forward matrix. This holds for any other perspective as well. 
The simultaneous solution with two (or more) perspectives can therefore be computed by performing the solution procedure on the matrix which is created by stacking the matrices for the two perspectives on the coefficient direction -- i.e., if one perspective has $n0$ output pixels and the other has $n1$ output pixels, the individual matrices will be $n0\times m$ and $n1\times m$ ($m$ is the number of coefficients of the source model): the combined problem matrix therefore has dimensions $(n0+n1)\times m$. Figure \ref{fig:multiperspective_tableau} shows this for the 2-perspective inversion with AIA and STEREO A. 

\begin{figure}
    \begin{equation}
        \left[\begin{array}{c}
            d^\mathrm{AIA}_0\\
            d^\mathrm{AIA}_1\\
            \dots\\
            d^\mathrm{AIA}_{n0-1}\\
            \hline
            d^\mathrm{STA}_0\\
            d^\mathrm{STA}_1\\
            \dots\\
            d^\mathrm{STA}_{n1-1}\\
        \end{array}\right]
        =
        \left[\begin{array}{cccc}
            a^{\mathrm{AIA}}_{0,0} & a^{\mathrm{AIA}}_{0,1} & \dots & a^{\mathrm{AIA}}_{0,m-1}\\
            a^{\mathrm{AIA}}_{1,0} & a^{\mathrm{AIA}}_{0,1} & \dots & a^{\mathrm{AIA}}_{0,m-1}\\
            \dots & \dots & \dots & \dots \\
            a^{\mathrm{AIA}}_{n0-1,0} & a^{\mathrm{AIA}}_{n0-1,1} & \dots & a^{\mathrm{AIA}}_{n0-1,m-1}\\
            \hline
            a^{\mathrm{STA}}_{0,0} & a^{\mathrm{STA}}_{0,1} & \dots & a^{\mathrm{STA}}_{0,m-1}\\
            a^{\mathrm{STA}}_{1,0} & a^{\mathrm{STA}}_{0,1} & \dots & a^{\mathrm{STA}}_{0,m-1}\\
            \dots & \dots & \dots & \dots \\
            a^{\mathrm{STA}}_{n1-1,0} & a^{\mathrm{STA}}_{n1-1,1} & \dots & a^{\mathrm{STA}}_{n1-1,m-1}\\
        \end{array}\right]
        \left[\begin{array}{c}
            c_0\\
            c_1\\
            \dots\\
            c_{m-1}\\
        \end{array}\right]
    \end{equation}
    \caption{Illustration of how forward matrices of multiple perspectives (in this case, from SDO and STEREO A) are combined to make a single forward matrix that is then used by the reconstruction solver. The components of the $c$ vector are the coefficients of the solution (2 per field-aligned region), and there are $m$ of them in total. There is only one solution vector in common across all observations. The components of the $d^\mathrm{AIA}, d^\mathrm{STA}$ vectors are the values observed at each pixel in AIA and STEREO A respectively. There are $n0+n1$ of them in all, where $n0$, and $n1$ are the number of pixels in AIA and STEREO A respectively. The forward matrices $a^\mathrm{AIA}$ and $a^\mathrm{STA}$ individually map from coefficients to the pixels in their respective images as previously described. They are then simply stacked on top of each other row-wise to produce the overall matrix. Solution of the forward problem then proceeds exactly as described in \cite{CROBAR_2021}.}\label{fig:multiperspective_tableau}
\end{figure}

\subsection{Passband question}
\label{sec:passband_question}

For these multi-instrument comparisons, we are limited by the available observations to passbands which don't exactly match the power-law temperature response functions (e.g., $\sim$$T^2$), described by \citet{CROBAR_2021}, for which the real emission best matches the linear coefficient model used by CROBAR. However, Figure~\ref{fig:AR_example_image} has shown that EUVI 284\,\AA\ and AIA 335\,\AA\ are nevertheless visually quite similar to this power-law. Moreover, it mathematically suffices for the emitting regions to have comparable profiles of emission vs. arc length, which appears to be the case for these passbands. We therefore believe that these passbands suffice for using CROBAR at this level of development. At a later stage, it should be possible to refine the emission profiles of the $\mathbf{B}$-aligned regions independently, resolving this shortcoming.

\subsection{Validation}
\label{sec:validation}
The time interval we have chosen is specifically to allow validation of CROBAR using real data: the reconstruction with AIA alone can be checked against STEREO A and STEREO B, the reconstruction with AIA+STEREO B can be checked against STEREO A, and the reconstruction with AIA+STEREO A can be checked against STEREO B. In each case, the reconstruction allows all vantage points to be synthesized, and therefore a $\chi^2$ can be computed directly.

Synthesized images from the reconstructions are shown in Figures~\ref{fig:stereoa_reconstructions}--\ref{fig:aia_stereo_reconstruction}. First, Figure \ref{fig:stereoa_reconstructions} shows the (STEREO) EUVI-A 284\,\AA\ images computed first from AIA alone, then from AIA in conjunction with EUVI-B 284\,{\AA\}; the actual STEREO A data is shown as well, for reference. Next, Figure \ref{fig:stereob_reconstructions} shows EUVI-B images computed from AIA alone and from AIA plus EUVI-A, along with the actual EUVI-B image. Last, AIA synthetic 284\,\AA\ reconstructed from the EUVI-A and -B images is shown in Figure~\ref{fig:aia_stereo_reconstruction}.

For these figures we have used the best $\alpha$ (3.5 Gm$^{-1}$) inferred from the AIA reconstructions alone, and used that skeleton as the basis for the multi-perspective reconstructions here. However, we can also optimize $\alpha$ using the residuals of any combination of the perspectives without much additional effort, and intend to do so in future work.

\begin{figure}
    \includegraphics[width=\textwidth]{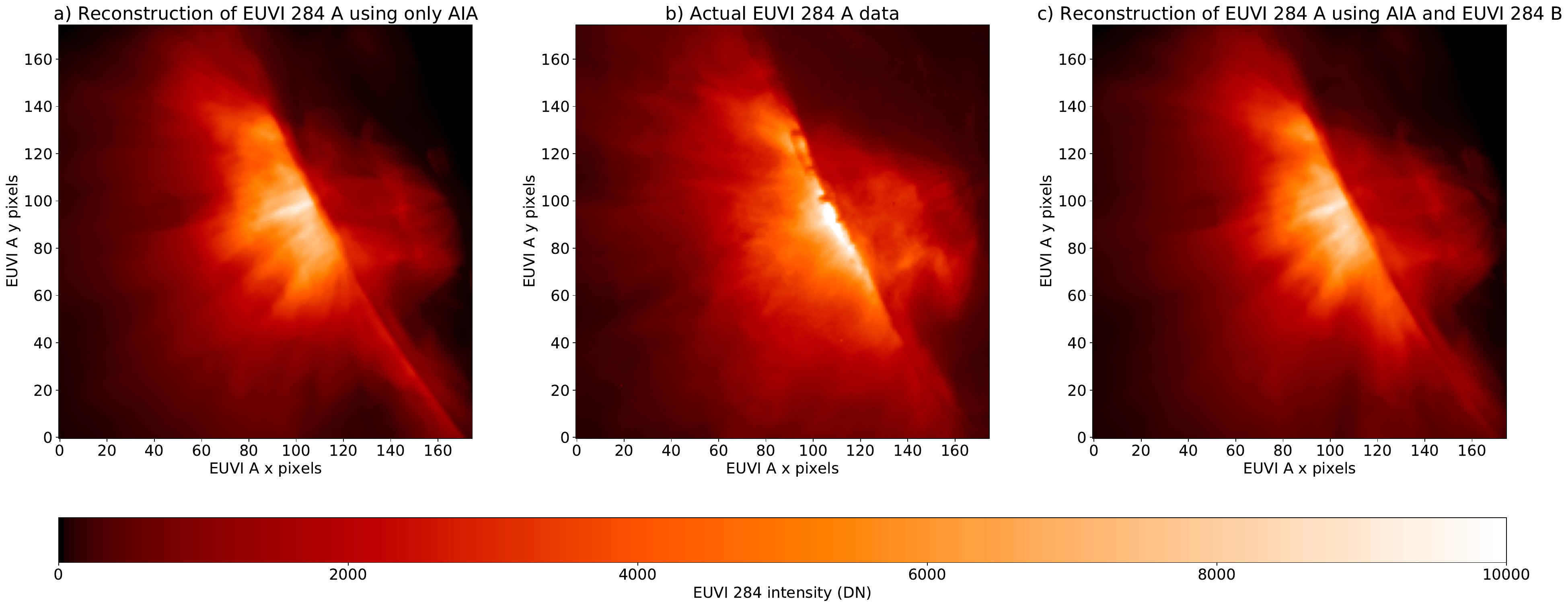}
    \caption{Demonstration and visual comparison of reconstruction of (STEREO) EUVI-A using AIA and EUVI-B. Left (a) shows reconstruction using AIA (284 \AA\ passband images synthesized using DEMs) alone, Center (b) shows the actual STEREO A data, Right (c) shows reconstruction using AIA and EUVI-B together. A significant goodness-of-fit improvement is seen when two perspectives are used instead of one, with typical error levels going from $\sim 33\%$ to $\sim 25\%$.}\label{fig:stereoa_reconstructions}
\end{figure}

\begin{figure}
    \includegraphics[width=\textwidth]{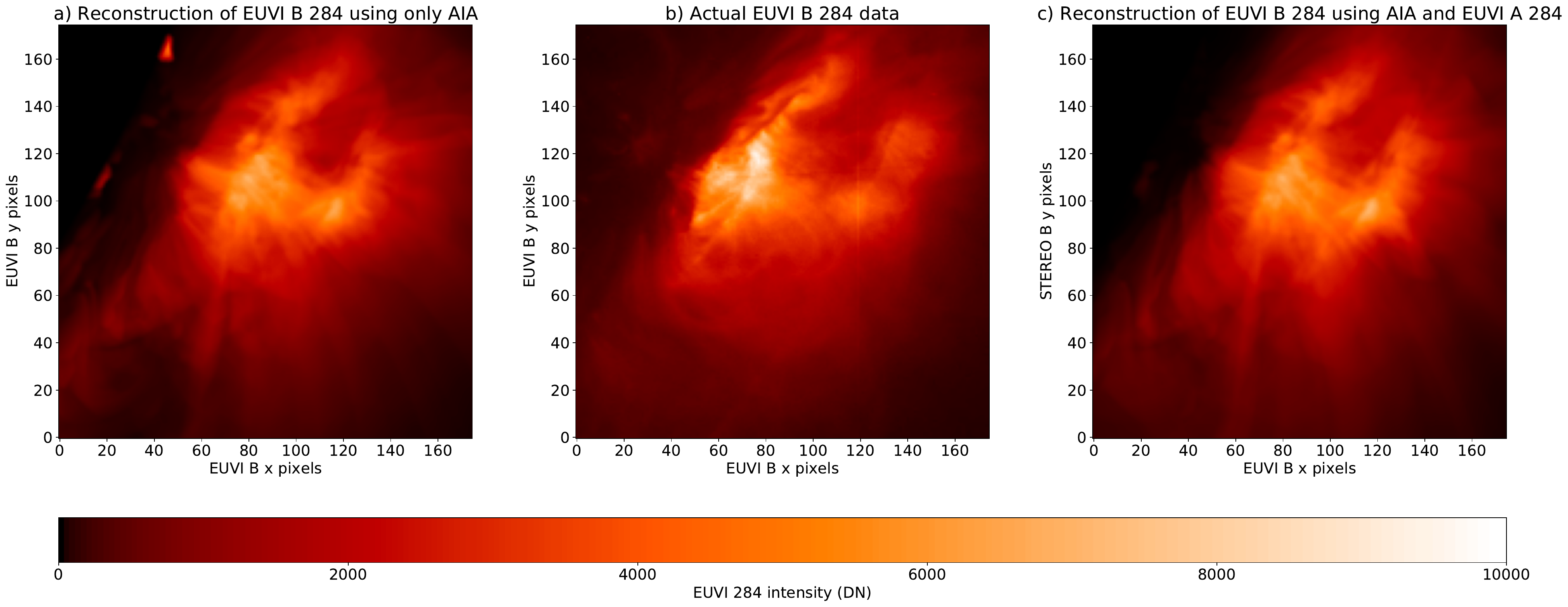}
    \caption{Demonstration and visual comparison of reconstruction of (STEREO) EUVI-B using AIA and STEREO A. Left (a) shows reconstruction using AIA (284 \AA\ passband images synthesized using DEMs) alone, Center (b) shows the actual EUVI-B data, Right (c) shows reconstruction using AIA and EUVI-A together. A significant goodness-of-fit improvement is seen when two perspectives are used instead of one, with typical error levels going from $\sim 33\%$ to $\sim 25\%$.}\label{fig:stereob_reconstructions}
\end{figure}

\begin{figure}
    \includegraphics[width=\textwidth]{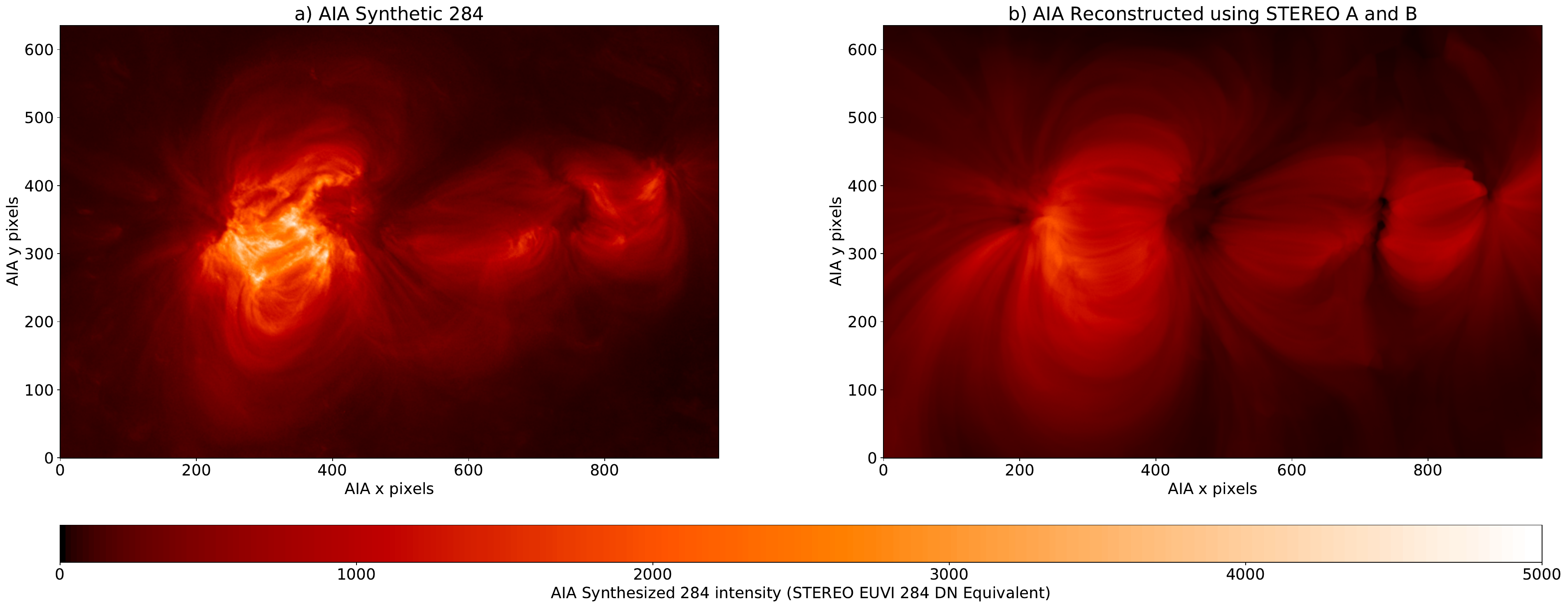}
    \caption{Demonstration and visual comparison of reconstruction of AIA using observations from the two STEREO EUVI spacecraft. Left (a) shows the original AIA data (synthesized 284 \AA\ observations made from DEMs), right (b) shows the reconstruction.}\label{fig:aia_stereo_reconstruction}
\end{figure}


For the $\chi^2$ that is the goodness-of-fit metric for the inversion, we assume a shot noise of 1~DN per photon plus 5\% for other systematic errors, plus a floor of 5~DN. The AIA-derived 284~{\AA\} images, based on the temperature response function shown in Figure \ref{fig:estimated_284_tresp}, are divided by an empirically estimated normalization factor of 4.8. This procedure results in roughly the same counts between the instruments. These values are then treated as DN for AIA, and the same error estimation procedure as for EUVI is applied. 
	
Our estimate of the actual quality of the reconstruction comes from comparing the reconstructions made with one and two vantage points by comparing with other vantage point (e.g., comparing the AIA+EUVI-A reconstruction with EUVI-B). For this comparison, we want to know an overall and more absolute level rather than whatever is set by the instrumental counts. We do still want to weight brighter regions higher, however. Therefore we used the same shot-noise plus relative plus floor estimate, and again computed a $\chi^2$, but experimented with the shot noise and relative levels until we obtained an effective reduced $\chi^2$ of order 1. We found that this happened when the shot noise and relative errors for the median intensity in the active region (roughly 2500 DN) combined to 25\%. In that case, the AIA-only $\chi^2$ was 1.76 for the reconstruction of EUVI-A and 1.91 for the reconstruction of EUVI-B. The reconstruction of EUVI-A from AIA and EUVI-B was 1.33 while the reconstruction of EUVI-B from AIA and EUVI-A was 0.98. All of these reconstructions of the STEREO data had an effective reduced $\chi^2$ of 1 or better when this error level was relaxed to 33\%. Therefore the two-perspective of EUVI-B was correct within 25\% in the active region core and that of EUVI-A (using EUVI-B and AIA) a little less close, and all are within 33\%.

The $\chi^2$ is better in the case of computing EUVI-B from AIA and EUVI-A rather than for EUVI-A from the other two perspectives, most likely because EUVI-B's perspective is closer to AIA's, whereas the EUVI-A is farther from the other two, being nearly at quadrature. This suggests that the quadrature (90 degree) perspectives are slightly better for the reconstructions than non-quadrature. The improved localization afforded by the second perspective is particularly striking in this case. We also find that the reconstructed STEREO images are cuspier (possessing sharp features at low heights in the corona) than the actual images, while the reconstructed AIA image is less cuspy than the actual. This suggests that the synthesized 284\,\AA\ passband from AIA has a larger contribution from low-lying cool (low corona or photosphere) emission than the STEREO images. Optical depth may play a role here, and/or the EUVI 284\,\AA\ instruments may be less sensitive to low-temperature emission than our estimated 284\,\AA\ temperature response functions.

\begin{figure}
    \includegraphics[width=\textwidth]{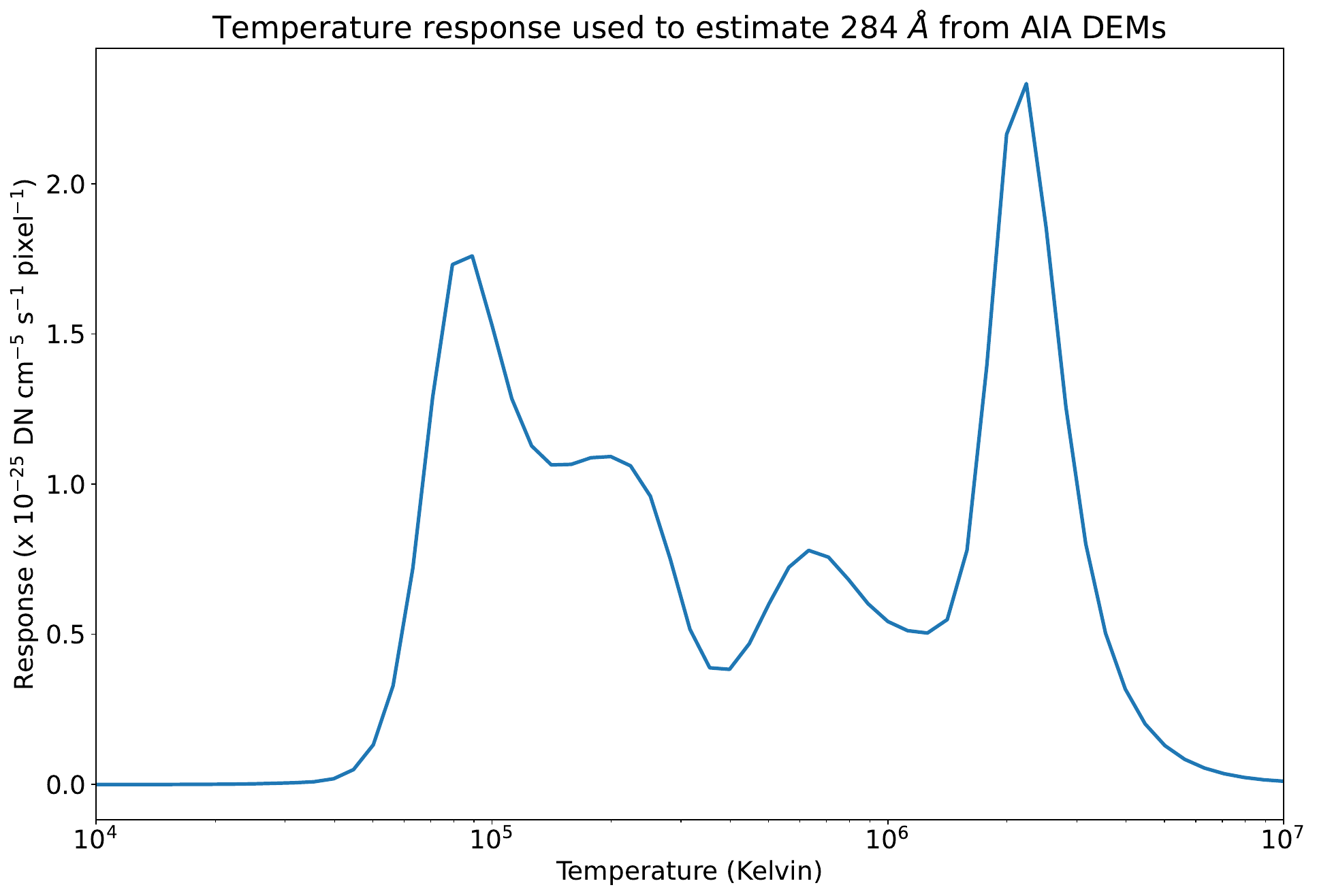}
    \caption{Temperature response function used to estimate the EUVI 284 \AA\ emission channel from AIA DEMs. An integration time of 1 second was assumed, and it was found that a relative normalization factor of 1/4.8 multiplied on this synthesized 284 \AA\ emission gave the best correspondence. The EUVI data used a 32 second exposure time, for comparison.}\label{fig:estimated_284_tresp}
\end{figure}


\section{Conclusions}
\label{sec:conclusion}
We have demonstrated both the ability of CROBAR to constrain the parameters of the solar magnetic field and that it can incorporate information from arbitrary perspectives, at arbitrary resolutions, into its reconstructions. The matrix-based treatment used in solving the forward problem allows highly heterogeneous data sources to be incorporated in a completely seamless fashion. CROBAR thus provides an invaluable starting point for true data-driven multi-instrument synthesis of the actual coronal volume. We have further shown how these reconstructions can give direct information on the physical parameters of the corona and the all-important magnetic energetics that drive the plasma. The results also indicate that emission channels such as 284\,\AA\ can have utility in its reconstructions even if they do not match the more ideal power law response functions. 

A 3D understanding of the corona is essential to truly transformative progress in unraveling many of the remaining mysteries in solar physics \citep{Caspi2023_science}. CROBAR provides a key first step to mapping from 2D remote sensing observables to a physics-based, data-constrained 3D environment. CROBAR or similar data-driven reconstructions will provide a new means to explore the 3D corona, and new and necessary approaches to data assimilation for advanced coronal models \citep{Seaton2023}. 

Unified 3D data assimilation and modeling frameworks will be foundational to the next generation of distributed solar observatories \citep{Raouafi2023, Gopalswamy2023} or for coordination across several missions with unique viewpoints \cite{Hassler2023}. Missions that leverage highly-complementary multi-perspective observations specifically to probe the deep 3D physics of the corona, such as the COMPLETE concept that combines global magnetography and broadband spectroscopic imaging \citep{Caspi2023_complete}, will \textit{require} tools like CROBAR to link their many observables in self-consistent physically meaningful ways.

We will make a preliminary version of CROBAR available for download along with publication of the paper. It will include an example notebook that performs a single-perspective reconstruction of AIA data using the 335\,\AA\ channel for demonstration purposes. 

\acknowledgements{The analysis in this work was carried out in Python, making use of NumPy \citep{numpy}, SciPy \citep{scipy}, AstroPy \citep{astropy, astropyII}, and SunPy \citep{sunpy}. This work was supported by NASA grant 80NSSC17K0598, by NASA under GSFC subcontract \# 80GSFC20C0053 to SwRI, and by SwRI Presidential Discretion Internal Research funding under project 19.R6225.
}


\bibliography{main}

\end{document}